\documentclass[aip, cha, reprint]{revtex4-1}

\usepackage{xspace,units}
\usepackage{graphicx, multirow}
\usepackage{amssymb, amsmath, amsfonts}
\usepackage{xcolor}
\usepackage{enumitem}
\usepackage{graphicx}
\usepackage{subfigure}
\usepackage{color}
\usepackage{units}

\newcommand{\var}{\ensuremath{\sigma^2}\xspace}
\newcommand{\lagone}{\ensuremath{\rho}\xspace}
\newcommand{\aroc}{\ensuremath{A_{\textrm{ROC}}}\xspace}
\newcommand{\meanpow}{\ensuremath{\overline{\mathcal{P}}}\xspace}
 
\begin{document}

\title{No evidence for critical slowing down prior to human epileptic seizures}

\author{Theresa Wilkat}
\affiliation{Department of Epileptology, University of Bonn Medical Centre, Venusberg-Campus 1, 53127 Bonn, Germany}
\affiliation{Helmholtz Institute for Radiation and Nuclear Physics, University of Bonn, Nussallee 14--16, 53115 Bonn, Germany}

\author{Thorsten Rings}
\affiliation{Department of Epileptology, University of Bonn Medical Centre, Venusberg-Campus 1, 53127 Bonn, Germany}
\affiliation{Helmholtz Institute for Radiation and Nuclear Physics, University of Bonn, Nussallee 14--16, 53115 Bonn, Germany}

\author{Klaus Lehnertz}
\email{klaus.lehnertz@ukbonn.de}
\affiliation{Department of Epileptology, University of Bonn Medical Centre, Venusberg-Campus 1, 53127 Bonn, Germany}
\affiliation{Helmholtz Institute for Radiation and Nuclear Physics, University of Bonn, Nussallee 14--16, 53115 Bonn, Germany}
\affiliation{Interdisciplinary Center for Complex Systems, University of Bonn, Br{\"u}hler Stra\ss{}e 7, 53175 Bonn, Germany}

\begin{abstract}
There is a ongoing debate whether generic early warning signals for critical transitions exist that can be applied across diverse systems.
The human epileptic brain is often considered as a prototypical system, given the devastating and, at times, even life-threatening nature of the extreme event epileptic seizure.
More than three decades of international effort has successfully identified predictors of imminent seizures. 
However, the suitability of typically applied early warning indicators for critical slowing down, namely variance and lag\nobreakdash-1 autocorrelation, for indexing seizure susceptibility is still controversially discussed.
Here, we investigated long-term, multichannel recordings of brain dynamics from 28 subjects with epilepsy.
Using a surrogate-based evaluation procedure of sensitivity and specificity of time-resolved estimates of early warning indicators, we found no evidence for critical slowing down prior to 105 epileptic seizures.
\end{abstract}
\maketitle
\frenchspacing

\begin{quotation}
A tipping point in a complex system is a threshold that, when exceeded, can lead to large and devastating changes in the state of the system.
If the system is getting closer to a tipping point, the restoration to its normal state after some perturbation
takes increasingly longer and is associated with an increase in the size of fluctuations.  
Both these aspects have been proposed as early warning signals of tipping, usually referred to as critical slowing down. 
Epileptic seizures have repeatedly been claimed as a potential field of application on constructing early warning signals through identifying characteristics of critical slowing down on the basis of electroencephalographic time series.
We investigate whether there is evidence for critical slowing down prior to seizures by investigating long-lasting, multichannel electroencephalographic recordings from 28 subjects with epilepsy.
Applying state-of-the-art statistical approaches that involve specifically designed surrogate tests, we do not find evidence for critical slowing down prior to more than 100 epileptic seizures.
\end{quotation}

\section{Introduction}
The phenomenon of a stable equilibrium losing its stability as a slowly varying control parameter or some external forcing passes some critical value (tipping point) is referred to as a critical transition~\cite{afrajmovich1994,thompson1994,dakos2008,scheffer2009,kuehn2011,lenton2012,sieber2012,scheffer2012,dai2012generic,feudel2018multistability,kaszas2019tipping}.
A critical transition can be heralded by the phenomenon of critical slowing down (CSD), an increasingly slow recovery from small perturbations. 
In order to test whether a system undergoes CSD, one would ideally investigate its response to such small perturbations, although such an approach may not be feasible in general.  
An alternatively, data-driven approach consists of interpreting fluctuations in the state of a system as its responds to perturbations~\cite{dakos2008,scheffer2009,scheffer2012,dai2012generic,dakos2012methods,lenton2012}. 
With approaching the tipping point, the time needed for a system to recover from perturbations becomes longer and hence its dynamics becomes more correlated with its past, leading to an increase in the lag-1 autocorrelation estimated from time series of appropriate system observables.
In addition, since perturbations accumulate --~following Kubo's fluctuation-dissipation theorem~\cite{kubo1966}~-- one observes an increase in the size of the fluctuations (variance, or other higher-order statistical indicators) in such time series.
These as well as other indicators for CSD (e.g., derived from Fourier-transformed time series of system observables~\cite{scheffer2012}) 
are related to each other via Wiener-Khinchin's theorem and Parseval's (or Plancherel's) theorem.
Although indicators were claimed to be generic early warning signals, they have been critically discussed from various perspectives~\cite{ditlevsen2010tipping,boettiger2012early,boettiger2013b,boettiger2013no,guttal2013,dakos2015,dai2015relation,diks2015critical,wagner2015false,zhang2015predictability,milanowski2016seizures,qin2018early,oregan2018stochasticity,romano2018analysis,wen2018one,clements2019early,jager2019systematically}.
\begin{figure*}[ht!]
\includegraphics{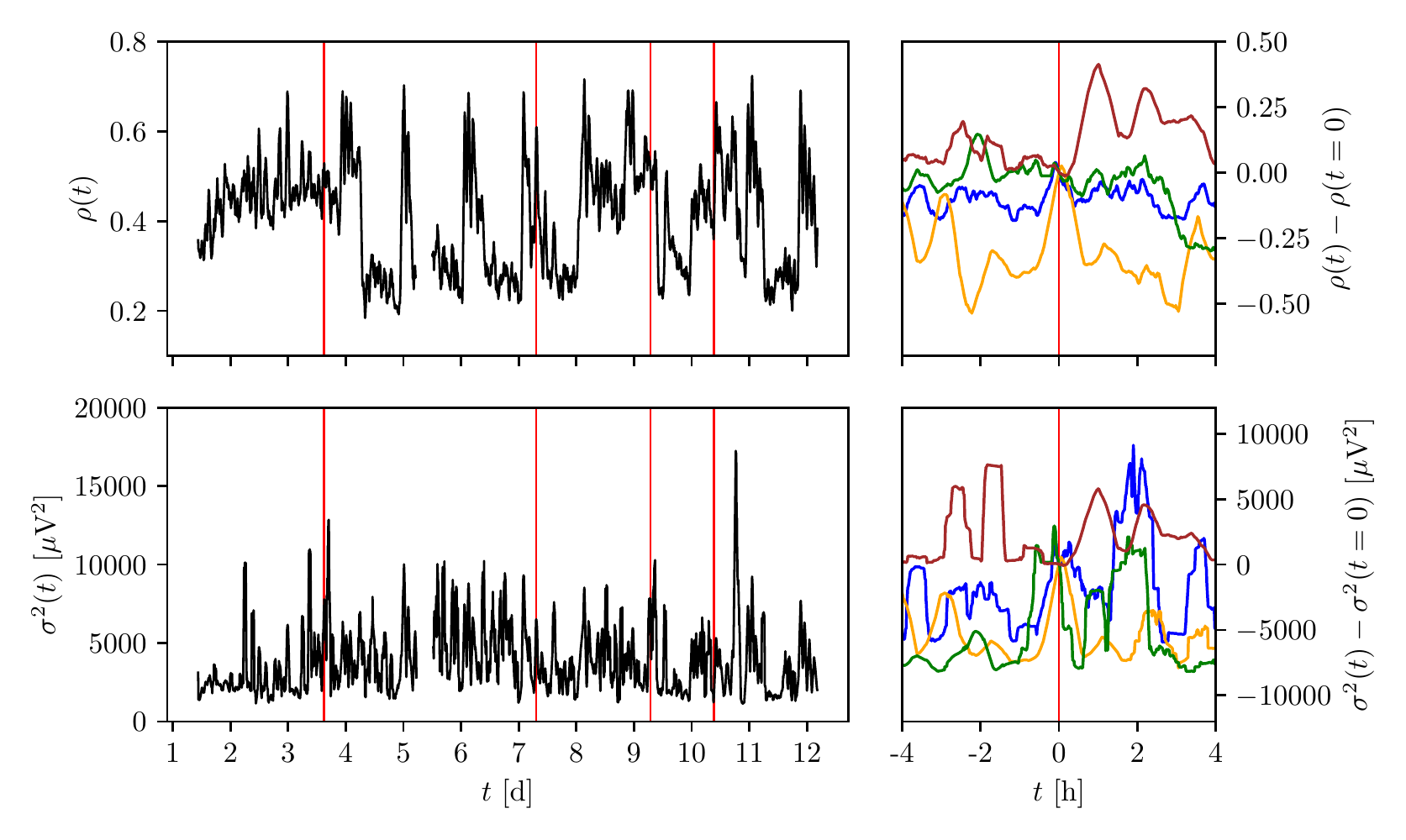}
	\caption{Left: exemplary time series of indicators of critical slowing down (top: \lagone; bottom: \var) from an iEEG recording during which four seizures were captured (seizure onsets marked with red vertical lines).
	To improve legibility, time series were smoothed with a moving average over 176 windows corresponding to \unit[60]{min}. 
	Discontinuities in the time series are due to recording gaps. 
	Tics on x-axes denote midnight.
	Right: zoom into time series of indicators when transiting into and out of epileptic seizures.
Indicator values of the pre- and post-seizure period ($\pm \unit[4]{h}$) are rescaled to the respective value at seizure onset
($t = 0$). 
Time series were smoothed with a moving average over 88	 windows corresponding to  \unit[30]{min}.
Different colors indicate different seizures.}
	\label{fig:fig1}
\end{figure*}

The human brain is among the many natural systems in which CSD has been repeatedly claimed to provide early warning signals for a variety of particularly pathophysiologic conditions~\cite{scheffer2009,kramer2012human,meisel2012scaling,leemput2014depression,meisel2015critical,bayani2017critical,scheffer2018quantifying}, most notably transitions into and out of epileptic seizures~\cite{scheffer2009,kramer2012human,meisel2012scaling,freestone2017forward,nazarimehr2018does}.
Previous studies~\cite{mormann2005,milanowski2016seizures}, however, could not identify clear cut indications for CSD in electroencephalographic data from the human epileptic brain. 
This discrepancy might have been caused by a data selection bias and/or the high variability of brain dynamics seen within and between subjects with epilepsy in between and prior to seizures~\cite{mormann2007,kuhlmann2018}.
Here we address this issue and investigate retrospectively long-term, multichannel intracranial electroencephalographic (iEEG) recordings from 28 subjects with epilepsy.
Using a surrogate-based evaluation procedure of sensitivity and specificity of time-resolved estimates of variance and lag\nobreakdash-1 autocorrelation, we found no evidence for CSD prior to 105 epileptic seizures.

\section{Data and Methods}
We investigated \unit[2056]{h} iEEG data recorded from about 1650 brain sites in 28 subjects with epilepsy who underwent pre-surgical evaluation at the Department of Epileptology of the University of Bonn.
These subjects suffered from pharmacoresistant focal seizures with different anatomical onset locations.
The study was approved by the ethics committee of the University of Bonn, and all patients had signed
informed consent that their clinical data might be used and published for research purposes. 

These multi-day, multi-channel data were part of previous studies~\cite{dickten2016,lehnertz2016,rings2019traceability,rings2019b}, were band-pass-filtered between \unit[1~--~45]{Hz}, sampled at \unit[200]{Hz} using a 16 bit analogue-to-digital converter, and were referenced against the average of two electrode contacts outside the presumed focal brain region. 
Reference contacts were chosen individually for each patient, and their data was disregarded in this study.
About 30\,\% of electrode contacts were confined to the brain region(s) from which seizures originate and its (their) direct vicinity, while the other contacts sampled brain dynamics in different lobes on the same and the opposite brain hemisphere. 
The majority of investigated brain sites investigated here is usually deemed unaffected by the epileptic process.

In order to probe for indications of CSD prior to seizures, we proceeded as follows:

First, we estimated --~for each subject~-- (unbiased sample) variance $\var_{j,w}$ and lag-1 autocorrelation $\lagone_{j,w}$ of iEEG data from each brain site $j$ ($j=1,\ldots,S$, where $S$ denotes the number of recording sites) in a sliding-window fashion~\cite{lehnertz2017capturing} ($w$ denotes the window number, and each window had a length of \unit[20.48]{s}).
For estimating the lag\nobreakdash-1 autocorrelation \lagone, we chose the smallest time-delay (here $\tau=\unit[30]{ms}$), for which the autocorrelation function $R(\tau)$ is not dominated by spurious correlations induced by the applied low-pass filter.
In order to not bias our analyses with effects stemming from occasional high-amplitude artifacts in the iEEG, we omitted the upper 0.5\,\% of all \var values.
In addition, we estimated the local derivatives of \lagone and  \var time series.

Second, we compared the distributions of values of indicators for CSD from an assumed pre-seizure period with the remaining data employing the receiver-operating-characteristics (ROC)~\cite{egan1975}. 
This allows us to assess the overall separability of distributions in terms of sensitivity and specificity of an indicator. 
We excluded, however, the \unit[60]{min} interval after the onset of a seizure in order to not bias our analyses with effects from the seizure and particularly from the post-seizure period~\cite{soblume2010}. 
Given that seizure generation is likely to take place over minutes to hours~\cite{mormann2007,kuhlmann2018}, we assumed a pre-seizure period to last for $T_{\rm pre} \in \unit[\left\{4, 2, 1, 0.5\right\}]{h}$.
Eventually, we calculated the area \aroc under the ROC curve relative to the case of identical distributions. 
The absolute value of \aroc is than confined to the interval $\left[0,0.5\right]$, and a positive/negative \aroc value corresponds to indicators for critical slowing down to be increased/decreased during the pre-seizure periods.

Third, we assessed the statistical validity of our findings by employing the concept of seizure time surrogates~\cite{andrzejak2003}.
We designed for each subject a total of 19 different surrogate sets of randomized seizure onset times by randomly permutating the intervals between seizures including the interval between the first seizure and the beginning of the recording.
An indication for CSD prior to seizures can be considered significant ($p < 0.05$), if $\aroc>0$ for original seizure times and if \aroc exceeds the maximum one obtained with 19 the seizure time surrogates.
\begin{figure}[htbp]
\includegraphics[width=0.5\textwidth]{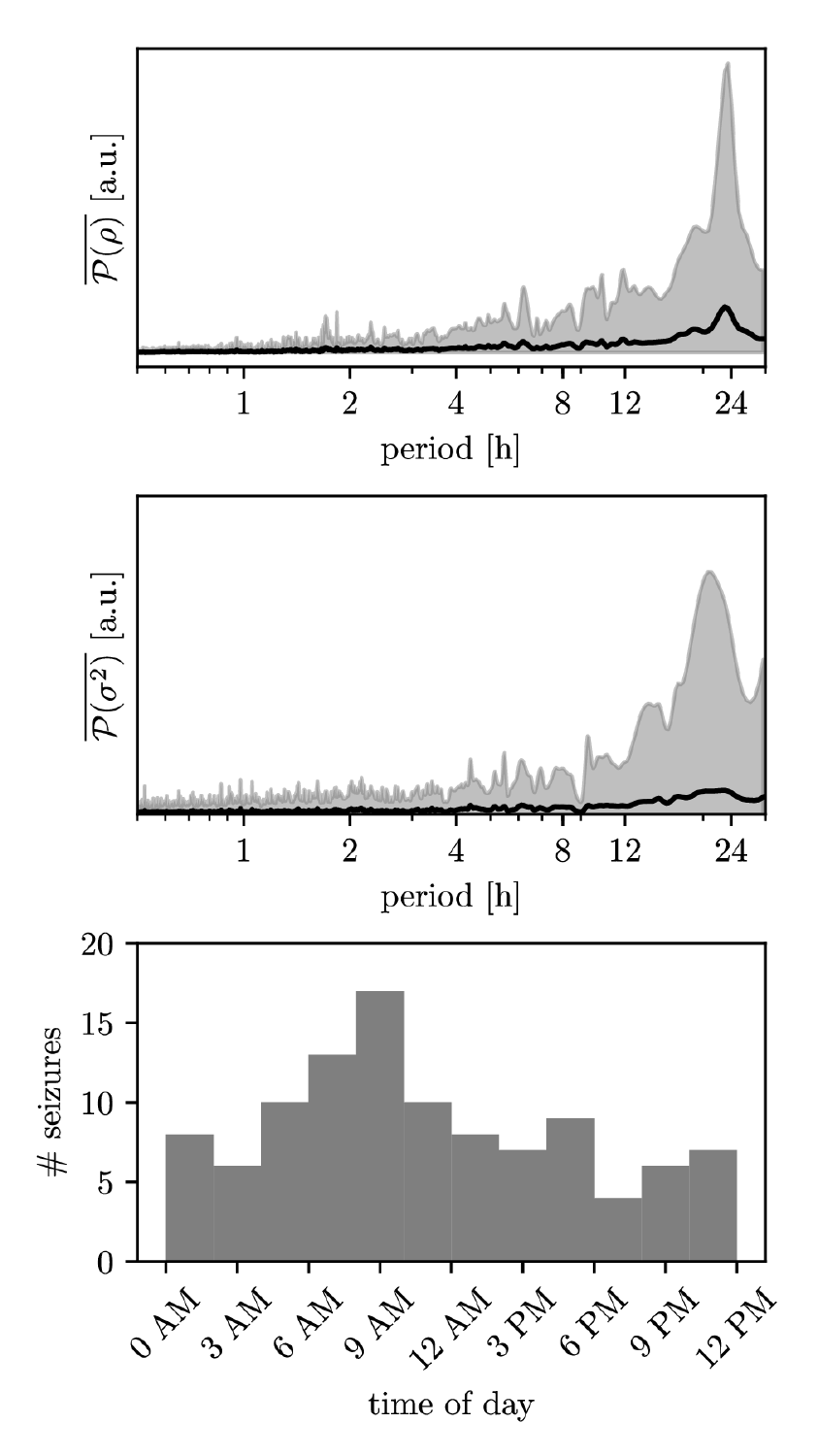}
	\caption{Top and middle: mean power spectral density estimates \meanpow of time series of indicators for critical slowing down \lagone and \var (mean over all recording sites) averaged over all patients. 
	Averaging was performed after interpolating (using cubic splines) the normalized power spectral density estimates (area under the curve equals 1) from individual patients in order to account for differences in sampling. 
	Mean values and ranges are shown as solid lines and shaded areas, respectively.
	Bottom: Distribution of occurrence of the 105 seizures investigated here over the \unit[24]{h} cycle.}
	\label{fig:fig2}
\end{figure}
\section{Results}
In the following, we present our findings for indicators lag\nobreakdash-1 autocorrelation \lagone and variance \var evaluated for pre-seizure periods lasting \unit[4]{h}. 
We obtained qualitatively similar findings for local derivatives of \lagone and  \var time series and for pre-seizure periods with shorter duration.
In Fig.~\ref{fig:fig1}, we show time series of indicators for CSD for an exemplary long-term recording covering twelve days.
If we consider a time window of \unit[4]{h} prior to seizures, we observe both indicators to either increase, to decrease, or to exhibit no clear-cut changes at all, on average.
We observe, however, a similar behavior at times far off epileptic seizures, which already puts into perspective considering  these indicators as generic early warning signals.

Interestingly, the dynamics of both indicators appears to be modulated on various timescales, most notably by daily rhythms such as the sleep-wake cycle. 
We therefore investigated the contribution of different timescales on the variability of indicators by estimating the power spectral density (Lomb-Scargle periodogram~\cite{press1989b}) of the unsmoothed time series of \var and \lagone from each electrode contact.
We observe power spectral densities for the investigated subjects to be qualitatively similar, with strong contributions at about \unit[24]{h} (Fig.~\ref{fig:fig2} top and middle). 
Processes acting on shorter timescales (\unit[\textless 4]{h}) contribute only marginally. 
Our findings indicate that a large fraction of the temporal variability of indicators for CSD can be attributed to processes acting on timescales of hours to days, with strong contributions of daily rhythms.

Currently, there is only weak evidence (mostly from modeling studies) for the wake-sleep/sleep-wake transition to represent a critical transition~\cite{steyn2005sleep,yang2016}.
Despite there being some studies that relate these physiologic transitions to the occurrence of seizures~\cite{stevens1971,karoly2016,baud2018,karoly2018}, the exact mechanisms underlying a possible relationship between sleep and epilepsy remain largely unknown~\cite{takagi2017sleep}.
We note that the seizures investigated here are not closely related to transitions into or out of sleep. 
Despite an increased incidence of seizures in the morning, which can be related to clinical investigations, seizures occurred equally distributed over the \unit[24]{h} cycle (Fig.~\ref{fig:fig2} bottom).
\begin{figure}[htbp]
\includegraphics[width=0.5\textwidth]{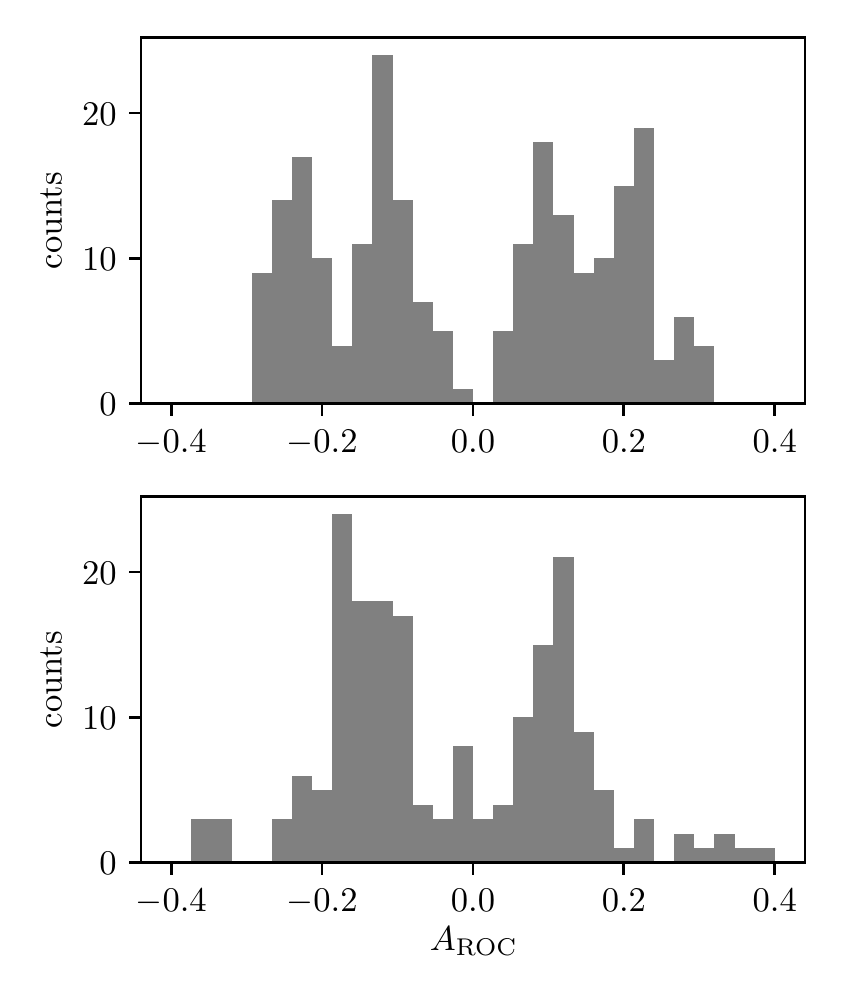}
\caption{Distribution of values of \aroc (area under the ROC curve) for CSD indicators \lagone (top) and \var (bottom) from brain sites that passed the surrogate test (235 sites using \lagone; 190 sites using \var).}
	\label{fig:fig3}
\end{figure}

In order to demonstrate extendability of these observations beyond exemplary data, we show for indicators \lagone and \var in Fig.~\ref{fig:fig3} the respective distributions of values of \aroc (area under the ROC curve) considered significant using our surrogate test. 
Of the 1647 brain sites, about one-seventh passed the test. 
From their corresponding \aroc values, we conclude that in the majority of cases both \lagone and \var rather point to a ``critical speeding up'' than to a critical slowing down.  
Although comparably high \aroc values can occasionally be observed, it is important to note that absolute values of \aroc for the original seizure times exceeded the maximum one of the seizure time surrogates by, on average, 7\,\%.
These findings indicate significant sensitivity and specificity of indicators for CSD for only a small number of brain sites.

Eventually, we investigated whether changes of indicators observed so far are spatially confined to the brain region(s) from which seizures originate. 
This was the case for 48 sites using \lagone, resp. 31 sites using \var, and there were indications for CSD at about half these sites (22 resp. 17). 
This could be traced back to the data from six subjects.
We checked whether these observations may result from the strongly varying number of electrode contacts that sample the various brain regions (hypergeometric tests; $p < 0.05$). 
This was the case for all six subjects when using \var as an indicator for CSD, and in four subjects when using \lagone.
In addition, a closer inspection of electrode types and locations in these subjects, however, revealed a spatial oversampling at about 50\,\% of these sites.

This leaves us with two to three subjects for which our chain of analysis indicates critical slowing down prior to seizures (based on lag\nobreakdash-1 autocorrelation) at about 1-2\% of their sampled brain sites, similar to what has been described before~\cite{mormann2005,milanowski2016seizures}.
We could not identify any peculiarities for these subjects.

\section{Conclusion}
Addressing a number of claims for critical slowing down prior to epileptic seizures, we investigated long-lasting, multichannel invasive electroencephalographic (iEEG) recordings that captured more than 100 seizures from 28 subjects with epilepsy.
We estimated the often-used early warning indicators for critical slowing down, namely variance and lag\nobreakdash-1 autocorrelation, from iEEG time series, and --~in contrast to many previous studies~-- utilized state-of-the-art statistical approaches to evaluate (temporal and spatial) sensitivity and specificity of indicators that involve specifically designed surrogate tests.

We found no clear-cut evidence for critical slowing down prior to epileptic seizures in humans.
We identified the sleep-wake cycle as a potential confounding variable --~among others~-- that might obscure detection of critical slowing down prior to seizures.
It is important to note that we lack clear-cut evidence for the wake-sleep/sleep-wake transition to represent a (physiologic) critical transition, and that the exact mechanisms underlying a possible sleep-induced occurrence of seizures remain largely unknown~\cite{takagi2017sleep,khan2018circadian}.

Also, we found no clear-cut evidence for critical slowing down to be confined to the brain region from which epileptic seizures appear to originate.
The lack of a spatial sensitivity and specificity of indicators for critical slowing down not only supports the current notion of 
a network mechanism to underlie the transition to the pre-seizure state~\cite{lehnertz2014,geier2015,lehnertz2016,kuhlmann2018,spencer2018roles,rings2019b} but also indicates that the mechanism behind the critical transition assumed here (bifurcation-induced tipping) may be too simplistic for the human epileptic brain.
There are other mechanisms behind tipping phenomena, such as noise-induced and rate-dependent tipping~\cite{ashwin2012tipping}.
Neither of these require any change of stability, and there may be no easy-identifiable early warning signals for such cases~\cite{ashwin2012tipping,ritchie2016,ritchie2017}. 

Research into seizure prediction~\cite{mormann2007,kuhlmann2018} has evaluated a large number of analysis concepts and methods to identify early warning signals for seizures, also with respect to clinical applicability. 
Univariate linear time series analysis techniques, such as the ones employed here, and also some univariate nonlinear techniques~\cite{mcsharry2003prediction} have been known to be insufficient for quite some time~\cite{mormann2005,mormann2007}.
We consider it a more promising approach to investigate how seizures emerge from large-scale brain networks taking into account their time-varying structure and function.

\section*{Acknowledgements}
We are grateful to Timo Br\"ohl for interesting discussions and for critical comments on earlier versions of the manuscript.


\begin{thebibliography}{65}%
\makeatletter
\providecommand \@ifxundefined [1]{%
 \@ifx{#1\undefined}
}%
\providecommand \@ifnum [1]{%
 \ifnum #1\expandafter \@firstoftwo
 \else \expandafter \@secondoftwo
 \fi
}%
\providecommand \@ifx [1]{%
 \ifx #1\expandafter \@firstoftwo
 \else \expandafter \@secondoftwo
 \fi
}%
\providecommand \natexlab [1]{#1}%
\providecommand \enquote  [1]{``#1''}%
\providecommand \bibnamefont  [1]{#1}%
\providecommand \bibfnamefont [1]{#1}%
\providecommand \citenamefont [1]{#1}%
\providecommand \href@noop [0]{\@secondoftwo}%
\providecommand \href [0]{\begingroup \@sanitize@url \@href}%
\providecommand \@href[1]{\@@startlink{#1}\@@href}%
\providecommand \@@href[1]{\endgroup#1\@@endlink}%
\providecommand \@sanitize@url [0]{\catcode `\\12\catcode `\$12\catcode
  `\&12\catcode `\#12\catcode `\^12\catcode `\_12\catcode `\%12\relax}%
\providecommand \@@startlink[1]{}%
\providecommand \@@endlink[0]{}%
\providecommand \url  [0]{\begingroup\@sanitize@url \@url }%
\providecommand \@url [1]{\endgroup\@href {#1}{\urlprefix }}%
\providecommand \urlprefix  [0]{URL }%
\providecommand \Eprint [0]{\href }%
\providecommand \doibase [0]{http://dx.doi.org/}%
\providecommand \selectlanguage [0]{\@gobble}%
\providecommand \bibinfo  [0]{\@secondoftwo}%
\providecommand \bibfield  [0]{\@secondoftwo}%
\providecommand \translation [1]{[#1]}%
\providecommand \BibitemOpen [0]{}%
\providecommand \bibitemStop [0]{}%
\providecommand \bibitemNoStop [0]{.\EOS\space}%
\providecommand \EOS [0]{\spacefactor3000\relax}%
\providecommand \BibitemShut  [1]{\csname bibitem#1\endcsname}%
\let\auto@bib@innerbib\@empty
\bibitem [{\citenamefont {Afrajmovich}\ \emph {et~al.}(1994)\citenamefont
  {Afrajmovich}, \citenamefont {Il'yashenko}, \citenamefont {Shil'nikov},
  \citenamefont {Arnold}, \citenamefont {Arnold},\ and\ \citenamefont
  {Kazarinoff}}]{afrajmovich1994}%
  \BibitemOpen
  \bibfield  {author} {\bibinfo {author} {\bibfnamefont {V.}~\bibnamefont
  {Afrajmovich}}, \bibinfo {author} {\bibfnamefont {Y.~S.}\ \bibnamefont
  {Il'yashenko}}, \bibinfo {author} {\bibfnamefont {L.}~\bibnamefont
  {Shil'nikov}}, \bibinfo {author} {\bibfnamefont {V.}~\bibnamefont {Arnold}},
  \bibinfo {author} {\bibfnamefont {V.}~\bibnamefont {Arnold}}, \ and\ \bibinfo
  {author} {\bibfnamefont {N.}~\bibnamefont {Kazarinoff}},\ }\href@noop {}
  {\emph {\bibinfo {title} {Dynamical Systems {V}: {B}ifurcation Theory and
  Catastrophe Theory}}}\ (\bibinfo  {publisher} {Springer},\ \bibinfo {year}
  {1994})\BibitemShut {NoStop}%
\bibitem [{\citenamefont {Thompson}, \citenamefont {Stewart},\ and\
  \citenamefont {Ueda}(1994)}]{thompson1994}%
  \BibitemOpen
  \bibfield  {author} {\bibinfo {author} {\bibfnamefont {J.~M.~T.}\
  \bibnamefont {Thompson}}, \bibinfo {author} {\bibfnamefont {H.~B.}\
  \bibnamefont {Stewart}}, \ and\ \bibinfo {author} {\bibfnamefont
  {Y.}~\bibnamefont {Ueda}},\ }\bibfield  {title} {\enquote {\bibinfo {title}
  {Safe, explosive, and dangerous bifurcations in dissipative dynamical
  systems},}\ }\href@noop {} {\bibfield  {journal} {\bibinfo  {journal} {Phys.
  Rev. E}\ }\textbf {\bibinfo {volume} {49}},\ \bibinfo {pages} {1019}
  (\bibinfo {year} {1994})}\BibitemShut {NoStop}%
\bibitem [{\citenamefont {Dakos}\ \emph {et~al.}(2008)\citenamefont {Dakos},
  \citenamefont {Scheffer}, \citenamefont {van Nes}, \citenamefont {Brovkin},
  \citenamefont {Petoukhov},\ and\ \citenamefont {Held}}]{dakos2008}%
  \BibitemOpen
  \bibfield  {author} {\bibinfo {author} {\bibfnamefont {V.}~\bibnamefont
  {Dakos}}, \bibinfo {author} {\bibfnamefont {M.}~\bibnamefont {Scheffer}},
  \bibinfo {author} {\bibfnamefont {E.~H.}\ \bibnamefont {van Nes}}, \bibinfo
  {author} {\bibfnamefont {V.}~\bibnamefont {Brovkin}}, \bibinfo {author}
  {\bibfnamefont {V.}~\bibnamefont {Petoukhov}}, \ and\ \bibinfo {author}
  {\bibfnamefont {H.}~\bibnamefont {Held}},\ }\bibfield  {title} {\enquote
  {\bibinfo {title} {Slowing down as an early warning signal for abrupt climate
  change},}\ }\href {\doibase 10.1073/pnas.0802430105} {\bibfield  {journal}
  {\bibinfo  {journal} {Proc. Natl. Acad. Sci. U.S.A.}\ }\textbf {\bibinfo
  {volume} {105}},\ \bibinfo {pages} {14308--14312} (\bibinfo {year}
  {2008})}\BibitemShut {NoStop}%
\bibitem [{\citenamefont {Scheffer}\ \emph {et~al.}(2009)\citenamefont
  {Scheffer}, \citenamefont {Bascompte}, \citenamefont {Brock}, \citenamefont
  {Brovkin}, \citenamefont {Carpenter}, \citenamefont {Dakos}, \citenamefont
  {Held}, \citenamefont {Van~Nes}, \citenamefont {Rietkerk},\ and\
  \citenamefont {Sugihara}}]{scheffer2009}%
  \BibitemOpen
  \bibfield  {author} {\bibinfo {author} {\bibfnamefont {M.}~\bibnamefont
  {Scheffer}}, \bibinfo {author} {\bibfnamefont {J.}~\bibnamefont {Bascompte}},
  \bibinfo {author} {\bibfnamefont {W.~A.}\ \bibnamefont {Brock}}, \bibinfo
  {author} {\bibfnamefont {V.}~\bibnamefont {Brovkin}}, \bibinfo {author}
  {\bibfnamefont {S.~R.}\ \bibnamefont {Carpenter}}, \bibinfo {author}
  {\bibfnamefont {V.}~\bibnamefont {Dakos}}, \bibinfo {author} {\bibfnamefont
  {H.}~\bibnamefont {Held}}, \bibinfo {author} {\bibfnamefont {E.~H.}\
  \bibnamefont {Van~Nes}}, \bibinfo {author} {\bibfnamefont {M.}~\bibnamefont
  {Rietkerk}}, \ and\ \bibinfo {author} {\bibfnamefont {G.}~\bibnamefont
  {Sugihara}},\ }\bibfield  {title} {\enquote {\bibinfo {title} {Early-warning
  signals for critical transitions},}\ }\href@noop {} {\bibfield  {journal}
  {\bibinfo  {journal} {Nature}\ }\textbf {\bibinfo {volume} {461}},\ \bibinfo
  {pages} {53} (\bibinfo {year} {2009})}\BibitemShut {NoStop}%
\bibitem [{\citenamefont {Kuehn}(2011)}]{kuehn2011}%
  \BibitemOpen
  \bibfield  {author} {\bibinfo {author} {\bibfnamefont {C.}~\bibnamefont
  {Kuehn}},\ }\bibfield  {title} {\enquote {\bibinfo {title} {A mathematical
  framework for critical transitions: Bifurcations, fast-slow systems and
  stochastic dynamics},}\ }\href {\doibase 10.1016/j.physd.2011.02.012}
  {\bibfield  {journal} {\bibinfo  {journal} {Physica~D}\ }\textbf {\bibinfo
  {volume} {240}},\ \bibinfo {pages} {1020--1035} (\bibinfo {year}
  {2011})}\BibitemShut {NoStop}%
\bibitem [{\citenamefont {Lenton}\ \emph {et~al.}(2012)\citenamefont {Lenton},
  \citenamefont {Livina}, \citenamefont {Dakos}, \citenamefont {Van~Nes},\ and\
  \citenamefont {Scheffer}}]{lenton2012}%
  \BibitemOpen
  \bibfield  {author} {\bibinfo {author} {\bibfnamefont {T.}~\bibnamefont
  {Lenton}}, \bibinfo {author} {\bibfnamefont {V.}~\bibnamefont {Livina}},
  \bibinfo {author} {\bibfnamefont {V.}~\bibnamefont {Dakos}}, \bibinfo
  {author} {\bibfnamefont {E.}~\bibnamefont {Van~Nes}}, \ and\ \bibinfo
  {author} {\bibfnamefont {M.}~\bibnamefont {Scheffer}},\ }\bibfield  {title}
  {\enquote {\bibinfo {title} {Early warning of climate tipping points from
  critical slowing down: comparing methods to improve robustness},}\
  }\href@noop {} {\bibfield  {journal} {\bibinfo  {journal} {Phil. Trans. Roy.
  Soc. A: Mathematical, Physical and Engineering Sciences}\ }\textbf {\bibinfo
  {volume} {370}},\ \bibinfo {pages} {1185--1204} (\bibinfo {year}
  {2012})}\BibitemShut {NoStop}%
\bibitem [{\citenamefont {Sieber}\ and\ \citenamefont
  {Thompson}(2012)}]{sieber2012}%
  \BibitemOpen
  \bibfield  {author} {\bibinfo {author} {\bibfnamefont {J.}~\bibnamefont
  {Sieber}}\ and\ \bibinfo {author} {\bibfnamefont {J.~M.~T.}\ \bibnamefont
  {Thompson}},\ }\bibfield  {title} {\enquote {\bibinfo {title} {Nonlinear
  softening as a predictive precursor to climate tipping},}\ }\href@noop {}
  {\bibfield  {journal} {\bibinfo  {journal} {Phil. Trans. Roy. Soc. A:
  Mathematical, Physical and Engineering Sciences}\ }\textbf {\bibinfo {volume}
  {370}},\ \bibinfo {pages} {1205--1227} (\bibinfo {year} {2012})}\BibitemShut
  {NoStop}%
\bibitem [{\citenamefont {Scheffer}\ \emph {et~al.}(2012)\citenamefont
  {Scheffer}, \citenamefont {Carpenter}, \citenamefont {Lenton}, \citenamefont
  {Bascompte}, \citenamefont {Brock}, \citenamefont {Dakos}, \citenamefont
  {van~de Koppel}, \citenamefont {van~de Leemput}, \citenamefont {Levin},
  \citenamefont {van Nes}, \citenamefont {Pascual},\ and\ \citenamefont
  {Vandermeer}}]{scheffer2012}%
  \BibitemOpen
  \bibfield  {author} {\bibinfo {author} {\bibfnamefont {M.}~\bibnamefont
  {Scheffer}}, \bibinfo {author} {\bibfnamefont {S.~R.}\ \bibnamefont
  {Carpenter}}, \bibinfo {author} {\bibfnamefont {T.~M.}\ \bibnamefont
  {Lenton}}, \bibinfo {author} {\bibfnamefont {J.}~\bibnamefont {Bascompte}},
  \bibinfo {author} {\bibfnamefont {W.}~\bibnamefont {Brock}}, \bibinfo
  {author} {\bibfnamefont {V.}~\bibnamefont {Dakos}}, \bibinfo {author}
  {\bibfnamefont {J.}~\bibnamefont {van~de Koppel}}, \bibinfo {author}
  {\bibfnamefont {I.~A.}\ \bibnamefont {van~de Leemput}}, \bibinfo {author}
  {\bibfnamefont {S.~A.}\ \bibnamefont {Levin}}, \bibinfo {author}
  {\bibfnamefont {E.~H.}\ \bibnamefont {van Nes}}, \bibinfo {author}
  {\bibfnamefont {M.}~\bibnamefont {Pascual}}, \ and\ \bibinfo {author}
  {\bibfnamefont {J.}~\bibnamefont {Vandermeer}},\ }\bibfield  {title}
  {\enquote {\bibinfo {title} {Anticipating critical transitions},}\ }\href
  {\doibase 10.1126/science.1225244} {\bibfield  {journal} {\bibinfo  {journal}
  {Science}\ }\textbf {\bibinfo {volume} {338}},\ \bibinfo {pages} {344--348}
  (\bibinfo {year} {2012})}\BibitemShut {NoStop}%
\bibitem [{\citenamefont {Dai}\ \emph {et~al.}(2012)\citenamefont {Dai},
  \citenamefont {Vorselen}, \citenamefont {Korolev},\ and\ \citenamefont
  {Gore}}]{dai2012generic}%
  \BibitemOpen
  \bibfield  {author} {\bibinfo {author} {\bibfnamefont {L.}~\bibnamefont
  {Dai}}, \bibinfo {author} {\bibfnamefont {D.}~\bibnamefont {Vorselen}},
  \bibinfo {author} {\bibfnamefont {K.~S.}\ \bibnamefont {Korolev}}, \ and\
  \bibinfo {author} {\bibfnamefont {J.}~\bibnamefont {Gore}},\ }\bibfield
  {title} {\enquote {\bibinfo {title} {Generic indicators for loss of
  resilience before a tipping point leading to population collapse},}\
  }\href@noop {} {\bibfield  {journal} {\bibinfo  {journal} {Science}\ }\textbf
  {\bibinfo {volume} {336}},\ \bibinfo {pages} {1175--1177} (\bibinfo {year}
  {2012})}\BibitemShut {NoStop}%
\bibitem [{\citenamefont {Feudel}, \citenamefont {Pisarchik},\ and\
  \citenamefont {Showalter}(2018)}]{feudel2018multistability}%
  \BibitemOpen
  \bibfield  {author} {\bibinfo {author} {\bibfnamefont {U.}~\bibnamefont
  {Feudel}}, \bibinfo {author} {\bibfnamefont {A.~N.}\ \bibnamefont
  {Pisarchik}}, \ and\ \bibinfo {author} {\bibfnamefont {K.}~\bibnamefont
  {Showalter}},\ }\bibfield  {title} {\enquote {\bibinfo {title}
  {Multistability and tipping: From mathematics and physics to climate and
  brain—minireview and preface to the focus issue},}\ }\href {\doibase
  10.1063/1.5027718} {\bibfield  {journal} {\bibinfo  {journal} {Chaos}\
  }\textbf {\bibinfo {volume} {28}},\ \bibinfo {pages} {033501} (\bibinfo
  {year} {2018})}\BibitemShut {NoStop}%
\bibitem [{\citenamefont {Kasz{\'a}s}, \citenamefont {Feudel},\ and\
  \citenamefont {T{\'e}l}(2019)}]{kaszas2019tipping}%
  \BibitemOpen
  \bibfield  {author} {\bibinfo {author} {\bibfnamefont {B.}~\bibnamefont
  {Kasz{\'a}s}}, \bibinfo {author} {\bibfnamefont {U.}~\bibnamefont {Feudel}},
  \ and\ \bibinfo {author} {\bibfnamefont {T.}~\bibnamefont {T{\'e}l}},\
  }\bibfield  {title} {\enquote {\bibinfo {title} {Tipping phenomena in typical
  dynamical systems subjected to parameter drift},}\ }\href@noop {} {\bibfield
  {journal} {\bibinfo  {journal} {Sci. Rep.}\ }\textbf {\bibinfo {volume}
  {9}},\ \bibinfo {pages} {8654} (\bibinfo {year} {2019})}\BibitemShut
  {NoStop}%
\bibitem [{\citenamefont {Dakos}\ \emph {et~al.}(2012)\citenamefont {Dakos},
  \citenamefont {Carpenter}, \citenamefont {Brock}, \citenamefont {Ellison},
  \citenamefont {Guttal}, \citenamefont {Ives}, \citenamefont {Kéfi},
  \citenamefont {Livina}, \citenamefont {Seekell}, \citenamefont {van Nes},\
  and\ \citenamefont {Scheffer}}]{dakos2012methods}%
  \BibitemOpen
  \bibfield  {author} {\bibinfo {author} {\bibfnamefont {V.}~\bibnamefont
  {Dakos}}, \bibinfo {author} {\bibfnamefont {S.~R.}\ \bibnamefont
  {Carpenter}}, \bibinfo {author} {\bibfnamefont {W.~A.}\ \bibnamefont
  {Brock}}, \bibinfo {author} {\bibfnamefont {A.~M.}\ \bibnamefont {Ellison}},
  \bibinfo {author} {\bibfnamefont {V.}~\bibnamefont {Guttal}}, \bibinfo
  {author} {\bibfnamefont {A.~R.}\ \bibnamefont {Ives}}, \bibinfo {author}
  {\bibfnamefont {S.}~\bibnamefont {Kéfi}}, \bibinfo {author} {\bibfnamefont
  {V.}~\bibnamefont {Livina}}, \bibinfo {author} {\bibfnamefont {D.~A.}\
  \bibnamefont {Seekell}}, \bibinfo {author} {\bibfnamefont {E.~H.}\
  \bibnamefont {van Nes}}, \ and\ \bibinfo {author} {\bibfnamefont
  {M.}~\bibnamefont {Scheffer}},\ }\bibfield  {title} {\enquote {\bibinfo
  {title} {Methods for detecting early warnings of critical transitions in time
  series illustrated using simulated ecological data},}\ }\href {\doibase
  10.1371/journal.pone.0041010} {\bibfield  {journal} {\bibinfo  {journal}
  {PLOS ONE}\ }\textbf {\bibinfo {volume} {7}},\ \bibinfo {pages} {1--20}
  (\bibinfo {year} {2012})}\BibitemShut {NoStop}%
\bibitem [{\citenamefont {Kubo}(1966)}]{kubo1966}%
  \BibitemOpen
  \bibfield  {author} {\bibinfo {author} {\bibfnamefont {R.}~\bibnamefont
  {Kubo}},\ }\bibfield  {title} {\enquote {\bibinfo {title} {The
  fluctuation-dissipation theorem},}\ }\href@noop {} {\bibfield  {journal}
  {\bibinfo  {journal} {Rep. Prog. Phys.}\ }\textbf {\bibinfo {volume} {29}},\
  \bibinfo {pages} {255} (\bibinfo {year} {1966})}\BibitemShut {NoStop}%
\bibitem [{\citenamefont {Ditlevsen}\ and\ \citenamefont
  {Johnsen}(2010)}]{ditlevsen2010tipping}%
  \BibitemOpen
  \bibfield  {author} {\bibinfo {author} {\bibfnamefont {P.~D.}\ \bibnamefont
  {Ditlevsen}}\ and\ \bibinfo {author} {\bibfnamefont {S.~J.}\ \bibnamefont
  {Johnsen}},\ }\bibfield  {title} {\enquote {\bibinfo {title} {Tipping points:
  early warning and wishful thinking},}\ }\href@noop {} {\bibfield  {journal}
  {\bibinfo  {journal} {Geophys. Res. Lett.}\ }\textbf {\bibinfo {volume}
  {37}},\ \bibinfo {pages} {L19703} (\bibinfo {year} {2010})}\BibitemShut
  {NoStop}%
\bibitem [{\citenamefont {Boettiger}\ and\ \citenamefont
  {Hastings}(2012)}]{boettiger2012early}%
  \BibitemOpen
  \bibfield  {author} {\bibinfo {author} {\bibfnamefont {C.}~\bibnamefont
  {Boettiger}}\ and\ \bibinfo {author} {\bibfnamefont {A.}~\bibnamefont
  {Hastings}},\ }\bibfield  {title} {\enquote {\bibinfo {title} {Early warning
  signals and the prosecutor's fallacy},}\ }\href@noop {} {\bibfield  {journal}
  {\bibinfo  {journal} {Proc Roy Soc. B: Biol. Sci.}\ }\textbf {\bibinfo
  {volume} {279}},\ \bibinfo {pages} {4734--4739} (\bibinfo {year}
  {2012})}\BibitemShut {NoStop}%
\bibitem [{\citenamefont {Boettiger}, \citenamefont {Ross},\ and\ \citenamefont
  {Hastings}(2013)}]{boettiger2013b}%
  \BibitemOpen
  \bibfield  {author} {\bibinfo {author} {\bibfnamefont {C.}~\bibnamefont
  {Boettiger}}, \bibinfo {author} {\bibfnamefont {N.}~\bibnamefont {Ross}}, \
  and\ \bibinfo {author} {\bibfnamefont {A.}~\bibnamefont {Hastings}},\
  }\bibfield  {title} {\enquote {\bibinfo {title} {Early warning signals: the
  charted and uncharted territories},}\ }\href {\doibase
  10.1007/s12080-013-0192-6} {\bibfield  {journal} {\bibinfo  {journal} {Theor.
  Ecol.}\ }\textbf {\bibinfo {volume} {6}},\ \bibinfo {pages} {255--264}
  (\bibinfo {year} {2013})}\BibitemShut {NoStop}%
\bibitem [{\citenamefont {Boettiger}\ and\ \citenamefont
  {Hastings}(2013)}]{boettiger2013no}%
  \BibitemOpen
  \bibfield  {author} {\bibinfo {author} {\bibfnamefont {C.}~\bibnamefont
  {Boettiger}}\ and\ \bibinfo {author} {\bibfnamefont {A.}~\bibnamefont
  {Hastings}},\ }\bibfield  {title} {\enquote {\bibinfo {title} {No early
  warning signals for stochastic transitions: insights from large deviation
  theory},}\ }\href@noop {} {\bibfield  {journal} {\bibinfo  {journal} {Proc.
  Roy. Soc. B: Biol. Sci.}\ }\textbf {\bibinfo {volume} {280}},\ \bibinfo
  {pages} {20131372} (\bibinfo {year} {2013})}\BibitemShut {NoStop}%
\bibitem [{\citenamefont {Guttal}, \citenamefont {Jayaprakash},\ and\
  \citenamefont {Tabbaa}(2013)}]{guttal2013}%
  \BibitemOpen
  \bibfield  {author} {\bibinfo {author} {\bibfnamefont {V.}~\bibnamefont
  {Guttal}}, \bibinfo {author} {\bibfnamefont {C.}~\bibnamefont {Jayaprakash}},
  \ and\ \bibinfo {author} {\bibfnamefont {O.~P.}\ \bibnamefont {Tabbaa}},\
  }\bibfield  {title} {\enquote {\bibinfo {title} {Robustness of early warning
  signals of regime shifts in time-delayed ecological models},}\ }\href
  {\doibase 10.1007/s12080-013-0194-4} {\bibfield  {journal} {\bibinfo
  {journal} {Theor. Ecol.}\ }\textbf {\bibinfo {volume} {6}},\ \bibinfo {pages}
  {271--283} (\bibinfo {year} {2013})}\BibitemShut {NoStop}%
\bibitem [{\citenamefont {Dakos}\ \emph {et~al.}(2015)\citenamefont {Dakos},
  \citenamefont {Carpenter}, \citenamefont {{van Nes}},\ and\ \citenamefont
  {Scheffer}}]{dakos2015}%
  \BibitemOpen
  \bibfield  {author} {\bibinfo {author} {\bibfnamefont {V.}~\bibnamefont
  {Dakos}}, \bibinfo {author} {\bibfnamefont {S.~R.}\ \bibnamefont
  {Carpenter}}, \bibinfo {author} {\bibfnamefont {E.~H.}\ \bibnamefont {{van
  Nes}}}, \ and\ \bibinfo {author} {\bibfnamefont {M.}~\bibnamefont
  {Scheffer}},\ }\bibfield  {title} {\enquote {\bibinfo {title} {Resilience
  indicators: prospects and limitations for early warnings of regime shifts},}\
  }\href {\doibase 10.1098/rstb.2013.0263} {\bibfield  {journal} {\bibinfo
  {journal} {Phil. Trans. R. Soc. B: Biol. Sci.}\ }\textbf {\bibinfo {volume}
  {370}},\ \bibinfo {pages} {20130263} (\bibinfo {year} {2015})}\BibitemShut
  {NoStop}%
\bibitem [{\citenamefont {Dai}, \citenamefont {Korolev},\ and\ \citenamefont
  {Gore}(2015)}]{dai2015relation}%
  \BibitemOpen
  \bibfield  {author} {\bibinfo {author} {\bibfnamefont {L.}~\bibnamefont
  {Dai}}, \bibinfo {author} {\bibfnamefont {K.~S.}\ \bibnamefont {Korolev}}, \
  and\ \bibinfo {author} {\bibfnamefont {J.}~\bibnamefont {Gore}},\ }\bibfield
  {title} {\enquote {\bibinfo {title} {Relation between stability and
  resilience determines the performance of early warning signals under
  different environmental drivers},}\ }\href@noop {} {\bibfield  {journal}
  {\bibinfo  {journal} {Proc. Natl. Acad. Sci. (U.S.A.)}\ }\textbf {\bibinfo
  {volume} {112}},\ \bibinfo {pages} {10056--10061} (\bibinfo {year}
  {2015})}\BibitemShut {NoStop}%
\bibitem [{\citenamefont {Diks}, \citenamefont {Hommes},\ and\ \citenamefont
  {Wang}(2015)}]{diks2015critical}%
  \BibitemOpen
  \bibfield  {author} {\bibinfo {author} {\bibfnamefont {C.}~\bibnamefont
  {Diks}}, \bibinfo {author} {\bibfnamefont {C.}~\bibnamefont {Hommes}}, \ and\
  \bibinfo {author} {\bibfnamefont {J.}~\bibnamefont {Wang}},\ }\bibfield
  {title} {\enquote {\bibinfo {title} {Critical slowing down as an early
  warning signal for financial crises?}}\ }\href {\doibase
  https://doi.org/10.1007/s00181-018-1527-3} {\bibfield  {journal} {\bibinfo
  {journal} {Empirical Economics}\ ,\ \bibinfo {pages} {1--28}} (\bibinfo
  {year} {2015})}\BibitemShut {NoStop}%
\bibitem [{\citenamefont {Wagner}\ and\ \citenamefont
  {Eisenman}(2015)}]{wagner2015false}%
  \BibitemOpen
  \bibfield  {author} {\bibinfo {author} {\bibfnamefont {T.~J.}\ \bibnamefont
  {Wagner}}\ and\ \bibinfo {author} {\bibfnamefont {I.}~\bibnamefont
  {Eisenman}},\ }\bibfield  {title} {\enquote {\bibinfo {title} {False alarms:
  How early warning signals falsely predict abrupt sea ice loss},}\ }\href@noop
  {} {\bibfield  {journal} {\bibinfo  {journal} {Geophys. Res. Lett.}\ }\textbf
  {\bibinfo {volume} {42}},\ \bibinfo {pages} {10--333} (\bibinfo {year}
  {2015})}\BibitemShut {NoStop}%
\bibitem [{\citenamefont {Zhang}, \citenamefont {Kuehn},\ and\ \citenamefont
  {Hallerberg}(2015)}]{zhang2015predictability}%
  \BibitemOpen
  \bibfield  {author} {\bibinfo {author} {\bibfnamefont {X.}~\bibnamefont
  {Zhang}}, \bibinfo {author} {\bibfnamefont {C.}~\bibnamefont {Kuehn}}, \ and\
  \bibinfo {author} {\bibfnamefont {S.}~\bibnamefont {Hallerberg}},\ }\bibfield
   {title} {\enquote {\bibinfo {title} {Predictability of critical
  transitions},}\ }\href {\doibase 10.1103/PhysRevE.92.052905} {\bibfield
  {journal} {\bibinfo  {journal} {Phys. Rev. E}\ }\textbf {\bibinfo {volume}
  {92}},\ \bibinfo {pages} {052905} (\bibinfo {year} {2015})}\BibitemShut
  {NoStop}%
\bibitem [{\citenamefont {Milanowski}\ and\ \citenamefont
  {Suffczynski}(2016)}]{milanowski2016seizures}%
  \BibitemOpen
  \bibfield  {author} {\bibinfo {author} {\bibfnamefont {P.}~\bibnamefont
  {Milanowski}}\ and\ \bibinfo {author} {\bibfnamefont {P.}~\bibnamefont
  {Suffczynski}},\ }\bibfield  {title} {\enquote {\bibinfo {title} {Seizures
  start without common signatures of critical transition},}\ }\href@noop {}
  {\bibfield  {journal} {\bibinfo  {journal} {Int. J. Neural Syst.}\ }\textbf
  {\bibinfo {volume} {26}},\ \bibinfo {pages} {1650053} (\bibinfo {year}
  {2016})}\BibitemShut {NoStop}%
\bibitem [{\citenamefont {Qin}\ and\ \citenamefont
  {Tang}(2018)}]{qin2018early}%
  \BibitemOpen
  \bibfield  {author} {\bibinfo {author} {\bibfnamefont {S.}~\bibnamefont
  {Qin}}\ and\ \bibinfo {author} {\bibfnamefont {C.}~\bibnamefont {Tang}},\
  }\bibfield  {title} {\enquote {\bibinfo {title} {Early-warning signals of
  critical transition: {E}ffect of extrinsic noise},}\ }\href@noop {}
  {\bibfield  {journal} {\bibinfo  {journal} {Phys. Rev. E}\ }\textbf {\bibinfo
  {volume} {97}},\ \bibinfo {pages} {032406} (\bibinfo {year}
  {2018})}\BibitemShut {NoStop}%
\bibitem [{\citenamefont {O'Regan}\ and\ \citenamefont
  {Burton}(2018)}]{oregan2018stochasticity}%
  \BibitemOpen
  \bibfield  {author} {\bibinfo {author} {\bibfnamefont {S.~M.}\ \bibnamefont
  {O'Regan}}\ and\ \bibinfo {author} {\bibfnamefont {D.~L.}\ \bibnamefont
  {Burton}},\ }\bibfield  {title} {\enquote {\bibinfo {title} {How
  stochasticity influences leading indicators of critical transitions},}\
  }\href@noop {} {\bibfield  {journal} {\bibinfo  {journal} {Bull. Math.
  Biol.}\ }\textbf {\bibinfo {volume} {80}},\ \bibinfo {pages} {1630--1654}
  (\bibinfo {year} {2018})}\BibitemShut {NoStop}%
\bibitem [{\citenamefont {Romano}\ and\ \citenamefont
  {Kuehn}(2018)}]{romano2018analysis}%
  \BibitemOpen
  \bibfield  {author} {\bibinfo {author} {\bibfnamefont {F.}~\bibnamefont
  {Romano}}\ and\ \bibinfo {author} {\bibfnamefont {C.}~\bibnamefont {Kuehn}},\
  }\bibfield  {title} {\enquote {\bibinfo {title} {Analysis and predictability
  of tipping points with leading-order nonlinear term},}\ }\href@noop {}
  {\bibfield  {journal} {\bibinfo  {journal} {Int. J. Bifurcation Chaos}\
  }\textbf {\bibinfo {volume} {28}},\ \bibinfo {pages} {1850103} (\bibinfo
  {year} {2018})}\BibitemShut {NoStop}%
\bibitem [{\citenamefont {Wen}, \citenamefont {Ciamarra},\ and\ \citenamefont
  {Cheong}(2018)}]{wen2018one}%
  \BibitemOpen
  \bibfield  {author} {\bibinfo {author} {\bibfnamefont {H.}~\bibnamefont
  {Wen}}, \bibinfo {author} {\bibfnamefont {M.~P.}\ \bibnamefont {Ciamarra}}, \
  and\ \bibinfo {author} {\bibfnamefont {S.~A.}\ \bibnamefont {Cheong}},\
  }\bibfield  {title} {\enquote {\bibinfo {title} {How one might miss early
  warning signals of critical transitions in time series data: A systematic
  study of two major currency pairs},}\ }\href@noop {} {\bibfield  {journal}
  {\bibinfo  {journal} {PloS one}\ }\textbf {\bibinfo {volume} {13}},\ \bibinfo
  {pages} {e0191439} (\bibinfo {year} {2018})}\BibitemShut {NoStop}%
\bibitem [{\citenamefont {Clements}, \citenamefont {McCarthy},\ and\
  \citenamefont {Blanchard}(2019)}]{clements2019early}%
  \BibitemOpen
  \bibfield  {author} {\bibinfo {author} {\bibfnamefont {C.~F.}\ \bibnamefont
  {Clements}}, \bibinfo {author} {\bibfnamefont {M.~A.}\ \bibnamefont
  {McCarthy}}, \ and\ \bibinfo {author} {\bibfnamefont {J.~L.}\ \bibnamefont
  {Blanchard}},\ }\bibfield  {title} {\enquote {\bibinfo {title} {Early warning
  signals of recovery in complex systems},}\ }\href@noop {} {\bibfield
  {journal} {\bibinfo  {journal} {Nat. Commun.}\ }\textbf {\bibinfo {volume}
  {10}},\ \bibinfo {pages} {1681} (\bibinfo {year} {2019})}\BibitemShut
  {NoStop}%
\bibitem [{\citenamefont {J{\"a}ger}\ and\ \citenamefont
  {F{\"u}llsack}(2019)}]{jager2019systematically}%
  \BibitemOpen
  \bibfield  {author} {\bibinfo {author} {\bibfnamefont {G.}~\bibnamefont
  {J{\"a}ger}}\ and\ \bibinfo {author} {\bibfnamefont {M.}~\bibnamefont
  {F{\"u}llsack}},\ }\bibfield  {title} {\enquote {\bibinfo {title}
  {Systematically false positives in early warning signal analysis},}\
  }\href@noop {} {\bibfield  {journal} {\bibinfo  {journal} {PloS one}\
  }\textbf {\bibinfo {volume} {14}},\ \bibinfo {pages} {e0211072} (\bibinfo
  {year} {2019})}\BibitemShut {NoStop}%
\bibitem [{\citenamefont {Kramer}\ \emph {et~al.}(2012)\citenamefont {Kramer},
  \citenamefont {Truccolo}, \citenamefont {Eden}, \citenamefont {Lepage},
  \citenamefont {Hochberg}, \citenamefont {Eskandar}, \citenamefont {Madsen},
  \citenamefont {Lee}, \citenamefont {Maheshwari}, \citenamefont {Halgren}
  \emph {et~al.}}]{kramer2012human}%
  \BibitemOpen
  \bibfield  {author} {\bibinfo {author} {\bibfnamefont {M.~A.}\ \bibnamefont
  {Kramer}}, \bibinfo {author} {\bibfnamefont {W.}~\bibnamefont {Truccolo}},
  \bibinfo {author} {\bibfnamefont {U.~T.}\ \bibnamefont {Eden}}, \bibinfo
  {author} {\bibfnamefont {K.~Q.}\ \bibnamefont {Lepage}}, \bibinfo {author}
  {\bibfnamefont {L.~R.}\ \bibnamefont {Hochberg}}, \bibinfo {author}
  {\bibfnamefont {E.~N.}\ \bibnamefont {Eskandar}}, \bibinfo {author}
  {\bibfnamefont {J.~R.}\ \bibnamefont {Madsen}}, \bibinfo {author}
  {\bibfnamefont {J.~W.}\ \bibnamefont {Lee}}, \bibinfo {author} {\bibfnamefont
  {A.}~\bibnamefont {Maheshwari}}, \bibinfo {author} {\bibfnamefont
  {E.}~\bibnamefont {Halgren}},  \emph {et~al.},\ }\bibfield  {title} {\enquote
  {\bibinfo {title} {Human seizures self-terminate across spatial scales via a
  critical transition},}\ }\href@noop {} {\bibfield  {journal} {\bibinfo
  {journal} {Proc. Natl. Acad. Sci. (U.S.A.)}\ }\textbf {\bibinfo {volume}
  {109}},\ \bibinfo {pages} {21116--21121} (\bibinfo {year}
  {2012})}\BibitemShut {NoStop}%
\bibitem [{\citenamefont {Meisel}\ and\ \citenamefont
  {Kuehn}(2012)}]{meisel2012scaling}%
  \BibitemOpen
  \bibfield  {author} {\bibinfo {author} {\bibfnamefont {C.}~\bibnamefont
  {Meisel}}\ and\ \bibinfo {author} {\bibfnamefont {C.}~\bibnamefont {Kuehn}},\
  }\bibfield  {title} {\enquote {\bibinfo {title} {Scaling effects and
  spatio-temporal multilevel dynamics in epileptic seizures},}\ }\href@noop {}
  {\bibfield  {journal} {\bibinfo  {journal} {PLoS One}\ }\textbf {\bibinfo
  {volume} {7}},\ \bibinfo {pages} {e30371} (\bibinfo {year}
  {2012})}\BibitemShut {NoStop}%
\bibitem [{\citenamefont {van~de Leemput}\ \emph {et~al.}(2014)\citenamefont
  {van~de Leemput}, \citenamefont {Wichers}, \citenamefont {Cramer},
  \citenamefont {Borsboom}, \citenamefont {Tuerlinckx}, \citenamefont
  {Kuppens}, \citenamefont {van Nes}, \citenamefont {Viechtbauer},
  \citenamefont {Giltay}, \citenamefont {Aggen} \emph
  {et~al.}}]{leemput2014depression}%
  \BibitemOpen
  \bibfield  {author} {\bibinfo {author} {\bibfnamefont {I.~A.}\ \bibnamefont
  {van~de Leemput}}, \bibinfo {author} {\bibfnamefont {M.}~\bibnamefont
  {Wichers}}, \bibinfo {author} {\bibfnamefont {A.~O.}\ \bibnamefont {Cramer}},
  \bibinfo {author} {\bibfnamefont {D.}~\bibnamefont {Borsboom}}, \bibinfo
  {author} {\bibfnamefont {F.}~\bibnamefont {Tuerlinckx}}, \bibinfo {author}
  {\bibfnamefont {P.}~\bibnamefont {Kuppens}}, \bibinfo {author} {\bibfnamefont
  {E.~H.}\ \bibnamefont {van Nes}}, \bibinfo {author} {\bibfnamefont
  {W.}~\bibnamefont {Viechtbauer}}, \bibinfo {author} {\bibfnamefont {E.~J.}\
  \bibnamefont {Giltay}}, \bibinfo {author} {\bibfnamefont {S.~H.}\
  \bibnamefont {Aggen}},  \emph {et~al.},\ }\bibfield  {title} {\enquote
  {\bibinfo {title} {Critical slowing down as early warning for the onset and
  termination of depression},}\ }\href@noop {} {\bibfield  {journal} {\bibinfo
  {journal} {Proc. Natl. Acad. Sci. (U.S.A.)}\ }\textbf {\bibinfo {volume}
  {111}},\ \bibinfo {pages} {87--92} (\bibinfo {year} {2014})}\BibitemShut
  {NoStop}%
\bibitem [{\citenamefont {Meisel}\ \emph {et~al.}(2015)\citenamefont {Meisel},
  \citenamefont {Klaus}, \citenamefont {Kuehn},\ and\ \citenamefont
  {Plenz}}]{meisel2015critical}%
  \BibitemOpen
  \bibfield  {author} {\bibinfo {author} {\bibfnamefont {C.}~\bibnamefont
  {Meisel}}, \bibinfo {author} {\bibfnamefont {A.}~\bibnamefont {Klaus}},
  \bibinfo {author} {\bibfnamefont {C.}~\bibnamefont {Kuehn}}, \ and\ \bibinfo
  {author} {\bibfnamefont {D.}~\bibnamefont {Plenz}},\ }\bibfield  {title}
  {\enquote {\bibinfo {title} {Critical slowing down governs the transition to
  neuron spiking},}\ }\href@noop {} {\bibfield  {journal} {\bibinfo  {journal}
  {PLoS Comput. Biol.}\ }\textbf {\bibinfo {volume} {11}},\ \bibinfo {pages}
  {e1004097} (\bibinfo {year} {2015})}\BibitemShut {NoStop}%
\bibitem [{\citenamefont {Bayani}\ \emph {et~al.}(2017)\citenamefont {Bayani},
  \citenamefont {Hadaeghi}, \citenamefont {Jafari},\ and\ \citenamefont
  {Murray}}]{bayani2017critical}%
  \BibitemOpen
  \bibfield  {author} {\bibinfo {author} {\bibfnamefont {A.}~\bibnamefont
  {Bayani}}, \bibinfo {author} {\bibfnamefont {F.}~\bibnamefont {Hadaeghi}},
  \bibinfo {author} {\bibfnamefont {S.}~\bibnamefont {Jafari}}, \ and\ \bibinfo
  {author} {\bibfnamefont {G.}~\bibnamefont {Murray}},\ }\bibfield  {title}
  {\enquote {\bibinfo {title} {Critical slowing down as an early warning of
  transitions in episodes of bipolar disorder: A simulation study based on a
  computational model of circadian activity rhythms},}\ }\href@noop {}
  {\bibfield  {journal} {\bibinfo  {journal} {Chronobiol. Int.}\ }\textbf
  {\bibinfo {volume} {34}},\ \bibinfo {pages} {235--245} (\bibinfo {year}
  {2017})}\BibitemShut {NoStop}%
\bibitem [{\citenamefont {Scheffer}\ \emph {et~al.}(2018)\citenamefont
  {Scheffer}, \citenamefont {Bolhuis}, \citenamefont {Borsboom}, \citenamefont
  {Buchman}, \citenamefont {Gijzel}, \citenamefont {Goulson}, \citenamefont
  {Kammenga}, \citenamefont {Kemp}, \citenamefont {van~de Leemput},
  \citenamefont {Levin}, \citenamefont {Martin}, \citenamefont {Melis},
  \citenamefont {{van Nes}}, \citenamefont {Romero},\ and\ \citenamefont {{Olde
  Rikkert}}}]{scheffer2018quantifying}%
  \BibitemOpen
  \bibfield  {author} {\bibinfo {author} {\bibfnamefont {M.}~\bibnamefont
  {Scheffer}}, \bibinfo {author} {\bibfnamefont {J.~E.}\ \bibnamefont
  {Bolhuis}}, \bibinfo {author} {\bibfnamefont {D.}~\bibnamefont {Borsboom}},
  \bibinfo {author} {\bibfnamefont {T.~G.}\ \bibnamefont {Buchman}}, \bibinfo
  {author} {\bibfnamefont {S.~M.}\ \bibnamefont {Gijzel}}, \bibinfo {author}
  {\bibfnamefont {D.}~\bibnamefont {Goulson}}, \bibinfo {author} {\bibfnamefont
  {J.~E.}\ \bibnamefont {Kammenga}}, \bibinfo {author} {\bibfnamefont
  {B.}~\bibnamefont {Kemp}}, \bibinfo {author} {\bibfnamefont {I.~A.}\
  \bibnamefont {van~de Leemput}}, \bibinfo {author} {\bibfnamefont
  {S.}~\bibnamefont {Levin}}, \bibinfo {author} {\bibfnamefont {C.~M.}\
  \bibnamefont {Martin}}, \bibinfo {author} {\bibfnamefont {R.~J.~F.}\
  \bibnamefont {Melis}}, \bibinfo {author} {\bibfnamefont {E.~H.}\ \bibnamefont
  {{van Nes}}}, \bibinfo {author} {\bibfnamefont {L.~M.}\ \bibnamefont
  {Romero}}, \ and\ \bibinfo {author} {\bibfnamefont {M.~G.~M.}\ \bibnamefont
  {{Olde Rikkert}}},\ }\bibfield  {title} {\enquote {\bibinfo {title}
  {Quantifying resilience of humans and other animals},}\ }\href@noop {}
  {\bibfield  {journal} {\bibinfo  {journal} {Proc. Natl. Acad. Sci. (U.S.A.)}\
  }\textbf {\bibinfo {volume} {115}},\ \bibinfo {pages} {11883--11890}
  (\bibinfo {year} {2018})}\BibitemShut {NoStop}%
\bibitem [{\citenamefont {Freestone}, \citenamefont {Karoly},\ and\
  \citenamefont {Cook}(2017)}]{freestone2017forward}%
  \BibitemOpen
  \bibfield  {author} {\bibinfo {author} {\bibfnamefont {D.~R.}\ \bibnamefont
  {Freestone}}, \bibinfo {author} {\bibfnamefont {P.~J.}\ \bibnamefont
  {Karoly}}, \ and\ \bibinfo {author} {\bibfnamefont {M.~J.}\ \bibnamefont
  {Cook}},\ }\bibfield  {title} {\enquote {\bibinfo {title} {A forward-looking
  review of seizure prediction},}\ }\href@noop {} {\bibfield  {journal}
  {\bibinfo  {journal} {Curr. Opin. Neurol.}\ }\textbf {\bibinfo {volume}
  {30}},\ \bibinfo {pages} {167--173} (\bibinfo {year} {2017})}\BibitemShut
  {NoStop}%
\bibitem [{\citenamefont {Nazarimehr}, \citenamefont {Golpayegani},\ and\
  \citenamefont {Hatef}(2018)}]{nazarimehr2018does}%
  \BibitemOpen
  \bibfield  {author} {\bibinfo {author} {\bibfnamefont {F.}~\bibnamefont
  {Nazarimehr}}, \bibinfo {author} {\bibfnamefont {S.~M. R.~H.}\ \bibnamefont
  {Golpayegani}}, \ and\ \bibinfo {author} {\bibfnamefont {B.}~\bibnamefont
  {Hatef}},\ }\bibfield  {title} {\enquote {\bibinfo {title} {Does the onset of
  epileptic seizure start from a bifurcation point?}}\ }\href@noop {}
  {\bibfield  {journal} {\bibinfo  {journal} {Eur. Phys. J. ST}\ }\textbf
  {\bibinfo {volume} {227}},\ \bibinfo {pages} {697--705} (\bibinfo {year}
  {2018})}\BibitemShut {NoStop}%
\bibitem [{\citenamefont {Mormann}\ \emph {et~al.}(2005)\citenamefont
  {Mormann}, \citenamefont {Kreuz}, \citenamefont {Rieke}, \citenamefont
  {Andrzejak}, \citenamefont {Kraskov}, \citenamefont {David}, \citenamefont
  {Elger},\ and\ \citenamefont {Lehnertz}}]{mormann2005}%
  \BibitemOpen
  \bibfield  {author} {\bibinfo {author} {\bibfnamefont {F.}~\bibnamefont
  {Mormann}}, \bibinfo {author} {\bibfnamefont {T.}~\bibnamefont {Kreuz}},
  \bibinfo {author} {\bibfnamefont {C.}~\bibnamefont {Rieke}}, \bibinfo
  {author} {\bibfnamefont {R.~G.}\ \bibnamefont {Andrzejak}}, \bibinfo {author}
  {\bibfnamefont {A.}~\bibnamefont {Kraskov}}, \bibinfo {author} {\bibfnamefont
  {P.}~\bibnamefont {David}}, \bibinfo {author} {\bibfnamefont {C.~E.}\
  \bibnamefont {Elger}}, \ and\ \bibinfo {author} {\bibfnamefont
  {K.}~\bibnamefont {Lehnertz}},\ }\bibfield  {title} {\enquote {\bibinfo
  {title} {On the predictability of epileptic seizures},}\ }\href {\doibase
  10.1016/j.clinph.2004.08.025} {\bibfield  {journal} {\bibinfo  {journal}
  {Clin. Neurophysiol.}\ }\textbf {\bibinfo {volume} {116}},\ \bibinfo {pages}
  {569--587} (\bibinfo {year} {2005})}\BibitemShut {NoStop}%
\bibitem [{\citenamefont {Mormann}\ \emph {et~al.}(2007)\citenamefont
  {Mormann}, \citenamefont {Andrzejak}, \citenamefont {Elger},\ and\
  \citenamefont {Lehnertz}}]{mormann2007}%
  \BibitemOpen
  \bibfield  {author} {\bibinfo {author} {\bibfnamefont {F.}~\bibnamefont
  {Mormann}}, \bibinfo {author} {\bibfnamefont {R.}~\bibnamefont {Andrzejak}},
  \bibinfo {author} {\bibfnamefont {C.~E.}\ \bibnamefont {Elger}}, \ and\
  \bibinfo {author} {\bibfnamefont {K.}~\bibnamefont {Lehnertz}},\ }\bibfield
  {title} {\enquote {\bibinfo {title} {Seizure prediction: the long and winding
  road},}\ }\href {\doibase 10.1093/brain/awl241} {\bibfield  {journal}
  {\bibinfo  {journal} {Brain}\ }\textbf {\bibinfo {volume} {130}},\ \bibinfo
  {pages} {314--333} (\bibinfo {year} {2007})}\BibitemShut {NoStop}%
\bibitem [{\citenamefont {Kuhlmann}\ \emph {et~al.}(2018)\citenamefont
  {Kuhlmann}, \citenamefont {Lehnertz}, \citenamefont {Richardson},
  \citenamefont {Schelter},\ and\ \citenamefont {Zaveri}}]{kuhlmann2018}%
  \BibitemOpen
  \bibfield  {author} {\bibinfo {author} {\bibfnamefont {L.}~\bibnamefont
  {Kuhlmann}}, \bibinfo {author} {\bibfnamefont {K.}~\bibnamefont {Lehnertz}},
  \bibinfo {author} {\bibfnamefont {M.~P.}\ \bibnamefont {Richardson}},
  \bibinfo {author} {\bibfnamefont {B.}~\bibnamefont {Schelter}}, \ and\
  \bibinfo {author} {\bibfnamefont {H.~P.}\ \bibnamefont {Zaveri}},\ }\bibfield
   {title} {\enquote {\bibinfo {title} {Seizure prediction---ready for a new
  era},}\ }\href@noop {} {\bibfield  {journal} {\bibinfo  {journal} {Nat Rev.
  Neurol.}\ ,\ \bibinfo {pages} {618--630}} (\bibinfo {year}
  {2018})}\BibitemShut {NoStop}%
\bibitem [{\citenamefont {Dickten}\ \emph {et~al.}(2016)\citenamefont
  {Dickten}, \citenamefont {Porz}, \citenamefont {Elger},\ and\ \citenamefont
  {Lehnertz}}]{dickten2016}%
  \BibitemOpen
  \bibfield  {author} {\bibinfo {author} {\bibfnamefont {H.}~\bibnamefont
  {Dickten}}, \bibinfo {author} {\bibfnamefont {S.}~\bibnamefont {Porz}},
  \bibinfo {author} {\bibfnamefont {C.~E.}\ \bibnamefont {Elger}}, \ and\
  \bibinfo {author} {\bibfnamefont {K.}~\bibnamefont {Lehnertz}},\ }\bibfield
  {title} {\enquote {\bibinfo {title} {Weighted and directed interactions in
  evolving large-scale epileptic brain networks},}\ }\href {\doibase
  10.1038/srep34824} {\bibfield  {journal} {\bibinfo  {journal} {Sci. Rep.}\
  }\textbf {\bibinfo {volume} {6}},\ \bibinfo {pages} {34824} (\bibinfo {year}
  {2016})}\BibitemShut {NoStop}%
\bibitem [{\citenamefont {Lehnertz}\ \emph {et~al.}(2016)\citenamefont
  {Lehnertz}, \citenamefont {Dickten}, \citenamefont {Porz}, \citenamefont
  {Helmstaedter},\ and\ \citenamefont {Elger}}]{lehnertz2016}%
  \BibitemOpen
  \bibfield  {author} {\bibinfo {author} {\bibfnamefont {K.}~\bibnamefont
  {Lehnertz}}, \bibinfo {author} {\bibfnamefont {H.}~\bibnamefont {Dickten}},
  \bibinfo {author} {\bibfnamefont {S.}~\bibnamefont {Porz}}, \bibinfo {author}
  {\bibfnamefont {C.}~\bibnamefont {Helmstaedter}}, \ and\ \bibinfo {author}
  {\bibfnamefont {C.~E.}\ \bibnamefont {Elger}},\ }\bibfield  {title} {\enquote
  {\bibinfo {title} {Predictability of uncontrollable multifocal seizures â€“
  towards new treatment options},}\ }\href {\doibase 10.1038/srep24584}
  {\bibfield  {journal} {\bibinfo  {journal} {Sci. Rep.}\ }\textbf {\bibinfo
  {volume} {6}},\ \bibinfo {pages} {24584} (\bibinfo {year}
  {2016})}\BibitemShut {NoStop}%
\bibitem [{\citenamefont {Rings}\ \emph {et~al.}(2019)\citenamefont {Rings},
  \citenamefont {Mazarei}, \citenamefont {Akhshi}, \citenamefont {Geier},
  \citenamefont {Tabar},\ and\ \citenamefont
  {Lehnertz}}]{rings2019traceability}%
  \BibitemOpen
  \bibfield  {author} {\bibinfo {author} {\bibfnamefont {T.}~\bibnamefont
  {Rings}}, \bibinfo {author} {\bibfnamefont {M.}~\bibnamefont {Mazarei}},
  \bibinfo {author} {\bibfnamefont {A.}~\bibnamefont {Akhshi}}, \bibinfo
  {author} {\bibfnamefont {C.}~\bibnamefont {Geier}}, \bibinfo {author}
  {\bibfnamefont {M.~R.~R.}\ \bibnamefont {Tabar}}, \ and\ \bibinfo {author}
  {\bibfnamefont {K.}~\bibnamefont {Lehnertz}},\ }\bibfield  {title} {\enquote
  {\bibinfo {title} {Traceability and dynamical resistance of precursor of
  extreme events},}\ }\href@noop {} {\bibfield  {journal} {\bibinfo  {journal}
  {Sci. Rep.}\ }\textbf {\bibinfo {volume} {9}},\ \bibinfo {pages} {1744}
  (\bibinfo {year} {2019})}\BibitemShut {NoStop}%
\bibitem [{\citenamefont {Rings}, \citenamefont {{von Wrede}},\ and\
  \citenamefont {Lehnertz}(2019)}]{rings2019b}%
  \BibitemOpen
  \bibfield  {author} {\bibinfo {author} {\bibfnamefont {T.}~\bibnamefont
  {Rings}}, \bibinfo {author} {\bibfnamefont {R.}~\bibnamefont {{von Wrede}}},
  \ and\ \bibinfo {author} {\bibfnamefont {K.}~\bibnamefont {Lehnertz}},\
  }\bibfield  {title} {\enquote {\bibinfo {title} {Precursors of seizures due
  to specific spatial-temporal modifications of evolving large-scale epileptic
  brain networks},}\ }\href@noop {} {\bibfield  {journal} {\bibinfo  {journal}
  {Sci. Rep.}\ }\textbf {\bibinfo {volume} {9}},\ \bibinfo {pages} {10623}
  (\bibinfo {year} {2019})}\BibitemShut {NoStop}%
\bibitem [{\citenamefont {Lehnertz}\ \emph {et~al.}(2017)\citenamefont
  {Lehnertz}, \citenamefont {Geier}, \citenamefont {Rings},\ and\ \citenamefont
  {Stahn}}]{lehnertz2017capturing}%
  \BibitemOpen
  \bibfield  {author} {\bibinfo {author} {\bibfnamefont {K.}~\bibnamefont
  {Lehnertz}}, \bibinfo {author} {\bibfnamefont {C.}~\bibnamefont {Geier}},
  \bibinfo {author} {\bibfnamefont {T.}~\bibnamefont {Rings}}, \ and\ \bibinfo
  {author} {\bibfnamefont {K.}~\bibnamefont {Stahn}},\ }\bibfield  {title}
  {\enquote {\bibinfo {title} {Capturing time-varying brain dynamics},}\
  }\href@noop {} {\bibfield  {journal} {\bibinfo  {journal} {EPJ Nonlin.
  Biomed. Phys.}\ }\textbf {\bibinfo {volume} {5}},\ \bibinfo {pages} {2}
  (\bibinfo {year} {2017})}\BibitemShut {NoStop}%
\bibitem [{\citenamefont {Egan}(1975)}]{egan1975}%
  \BibitemOpen
  \bibfield  {author} {\bibinfo {author} {\bibfnamefont {J.~P.}\ \bibnamefont
  {Egan}},\ }\href@noop {} {\emph {\bibinfo {title} {Signal detection theory
  and {ROC}-analysis}}}\ (\bibinfo  {publisher} {Academic press},\ \bibinfo
  {year} {1975})\BibitemShut {NoStop}%
\bibitem [{\citenamefont {So}\ and\ \citenamefont {Blume}(2010)}]{soblume2010}%
  \BibitemOpen
  \bibfield  {author} {\bibinfo {author} {\bibfnamefont {N.~K.}\ \bibnamefont
  {So}}\ and\ \bibinfo {author} {\bibfnamefont {W.~T.}\ \bibnamefont {Blume}},\
  }\bibfield  {title} {\enquote {\bibinfo {title} {The postictal {EEG}},}\
  }\href@noop {} {\bibfield  {journal} {\bibinfo  {journal} {Epilepsy Behav.}\
  }\textbf {\bibinfo {volume} {19}},\ \bibinfo {pages} {121--126} (\bibinfo
  {year} {2010})}\BibitemShut {NoStop}%
\bibitem [{\citenamefont {Andrzejak}\ \emph {et~al.}(2003)\citenamefont
  {Andrzejak}, \citenamefont {Mormann}, \citenamefont {Kreuz}, \citenamefont
  {Rieke}, \citenamefont {Kraskov}, \citenamefont {Elger},\ and\ \citenamefont
  {Lehnertz}}]{andrzejak2003}%
  \BibitemOpen
  \bibfield  {author} {\bibinfo {author} {\bibfnamefont {R.~G.}\ \bibnamefont
  {Andrzejak}}, \bibinfo {author} {\bibfnamefont {F.}~\bibnamefont {Mormann}},
  \bibinfo {author} {\bibfnamefont {T.}~\bibnamefont {Kreuz}}, \bibinfo
  {author} {\bibfnamefont {C.}~\bibnamefont {Rieke}}, \bibinfo {author}
  {\bibfnamefont {A.}~\bibnamefont {Kraskov}}, \bibinfo {author} {\bibfnamefont
  {C.~E.}\ \bibnamefont {Elger}}, \ and\ \bibinfo {author} {\bibfnamefont
  {K.}~\bibnamefont {Lehnertz}},\ }\bibfield  {title} {\enquote {\bibinfo
  {title} {Testing the null hypothesis of the nonexistence of a preseizure
  state},}\ }\href@noop {} {\bibfield  {journal} {\bibinfo  {journal} {Phys.
  Rev.~E}\ }\textbf {\bibinfo {volume} {67}},\ \bibinfo {pages} {010901(R)}
  (\bibinfo {year} {2003})}\BibitemShut {NoStop}%
\bibitem [{\citenamefont {Press}\ and\ \citenamefont
  {Rybicki}(1989)}]{press1989b}%
  \BibitemOpen
  \bibfield  {author} {\bibinfo {author} {\bibfnamefont {W.~H.}\ \bibnamefont
  {Press}}\ and\ \bibinfo {author} {\bibfnamefont {G.~B.}\ \bibnamefont
  {Rybicki}},\ }\bibfield  {title} {\enquote {\bibinfo {title} {Fast algorithm
  for spectral analysis of unevenly sampled data},}\ }\href {\doibase
  10.1086/167197} {\bibfield  {journal} {\bibinfo  {journal} {Astrophys.~J.}\
  }\textbf {\bibinfo {volume} {338}},\ \bibinfo {pages} {277--280} (\bibinfo
  {year} {1989})}\BibitemShut {NoStop}%
\bibitem [{\citenamefont {Steyn-Ross}\ \emph {et~al.}(2005)\citenamefont
  {Steyn-Ross}, \citenamefont {Steyn-Ross}, \citenamefont {Sleigh},
  \citenamefont {Wilson}, \citenamefont {Gillies},\ and\ \citenamefont
  {Wright}}]{steyn2005sleep}%
  \BibitemOpen
  \bibfield  {author} {\bibinfo {author} {\bibfnamefont {D.~A.}\ \bibnamefont
  {Steyn-Ross}}, \bibinfo {author} {\bibfnamefont {M.~L.}\ \bibnamefont
  {Steyn-Ross}}, \bibinfo {author} {\bibfnamefont {J.~W.}\ \bibnamefont
  {Sleigh}}, \bibinfo {author} {\bibfnamefont {M.~T.}\ \bibnamefont {Wilson}},
  \bibinfo {author} {\bibfnamefont {I.-P.}\ \bibnamefont {Gillies}}, \ and\
  \bibinfo {author} {\bibfnamefont {J.~J.}\ \bibnamefont {Wright}},\ }\bibfield
   {title} {\enquote {\bibinfo {title} {The sleep cycle modelled as a cortical
  phase transition},}\ }\href@noop {} {\bibfield  {journal} {\bibinfo
  {journal} {J. Biol. Phys.}\ }\textbf {\bibinfo {volume} {31}},\ \bibinfo
  {pages} {547--569} (\bibinfo {year} {2005})}\BibitemShut {NoStop}%
\bibitem [{\citenamefont {Yang}\ \emph {et~al.}(2016)\citenamefont {Yang},
  \citenamefont {McKenzie-Sell}, \citenamefont {Karanjai},\ and\ \citenamefont
  {Robinson}}]{yang2016}%
  \BibitemOpen
  \bibfield  {author} {\bibinfo {author} {\bibfnamefont {D.-P.}\ \bibnamefont
  {Yang}}, \bibinfo {author} {\bibfnamefont {L.}~\bibnamefont {McKenzie-Sell}},
  \bibinfo {author} {\bibfnamefont {A.}~\bibnamefont {Karanjai}}, \ and\
  \bibinfo {author} {\bibfnamefont {P.~A.}\ \bibnamefont {Robinson}},\
  }\bibfield  {title} {\enquote {\bibinfo {title} {Wake-sleep transition as a
  noisy bifurcation},}\ }\href {\doibase 10.1103/PhysRevE.94.022412} {\bibfield
   {journal} {\bibinfo  {journal} {Phys. Rev. E}\ }\textbf {\bibinfo {volume}
  {94}},\ \bibinfo {pages} {022412} (\bibinfo {year} {2016})}\BibitemShut
  {NoStop}%
\bibitem [{\citenamefont {Stevens}\ \emph {et~al.}(1971)\citenamefont
  {Stevens}, \citenamefont {Kodama}, \citenamefont {Lonsbury},\ and\
  \citenamefont {Mills}}]{stevens1971}%
  \BibitemOpen
  \bibfield  {author} {\bibinfo {author} {\bibfnamefont {J.}~\bibnamefont
  {Stevens}}, \bibinfo {author} {\bibfnamefont {H.}~\bibnamefont {Kodama}},
  \bibinfo {author} {\bibfnamefont {B.}~\bibnamefont {Lonsbury}}, \ and\
  \bibinfo {author} {\bibfnamefont {L.}~\bibnamefont {Mills}},\ }\bibfield
  {title} {\enquote {\bibinfo {title} {Ultradian characteristics of spontaneous
  seizures discharges recorded by radio telemetry in man},}\ }\href@noop {}
  {\bibfield  {journal} {\bibinfo  {journal} {Electroencephalogr. Clin.
  Neurophysiol.}\ }\textbf {\bibinfo {volume} {31}},\ \bibinfo {pages}
  {313--325} (\bibinfo {year} {1971})}\BibitemShut {NoStop}%
\bibitem [{\citenamefont {Karoly}\ \emph {et~al.}(2016)\citenamefont {Karoly},
  \citenamefont {Freestone}, \citenamefont {Boston}, \citenamefont {Grayden},
  \citenamefont {Himes}, \citenamefont {Leyde}, \citenamefont {Seneviratne},
  \citenamefont {Berkovic}, \citenamefont {O’Brien},\ and\ \citenamefont
  {Cook}}]{karoly2016}%
  \BibitemOpen
  \bibfield  {author} {\bibinfo {author} {\bibfnamefont {P.~J.}\ \bibnamefont
  {Karoly}}, \bibinfo {author} {\bibfnamefont {D.~R.}\ \bibnamefont
  {Freestone}}, \bibinfo {author} {\bibfnamefont {R.}~\bibnamefont {Boston}},
  \bibinfo {author} {\bibfnamefont {D.~B.}\ \bibnamefont {Grayden}}, \bibinfo
  {author} {\bibfnamefont {D.}~\bibnamefont {Himes}}, \bibinfo {author}
  {\bibfnamefont {K.}~\bibnamefont {Leyde}}, \bibinfo {author} {\bibfnamefont
  {U.}~\bibnamefont {Seneviratne}}, \bibinfo {author} {\bibfnamefont
  {S.}~\bibnamefont {Berkovic}}, \bibinfo {author} {\bibfnamefont
  {T.}~\bibnamefont {O’Brien}}, \ and\ \bibinfo {author} {\bibfnamefont
  {M.~J.}\ \bibnamefont {Cook}},\ }\bibfield  {title} {\enquote {\bibinfo
  {title} {Interictal spikes and epileptic seizures: their relationship and
  underlying rhythmicity},}\ }\href@noop {} {\bibfield  {journal} {\bibinfo
  {journal} {Brain}\ }\textbf {\bibinfo {volume} {139}},\ \bibinfo {pages}
  {1066--1078} (\bibinfo {year} {2016})}\BibitemShut {NoStop}%
\bibitem [{\citenamefont {Baud}\ \emph {et~al.}(2018)\citenamefont {Baud},
  \citenamefont {Kleen}, \citenamefont {Mirro}, \citenamefont {Andrechak},
  \citenamefont {King-Stephens}, \citenamefont {Chang},\ and\ \citenamefont
  {Rao}}]{baud2018}%
  \BibitemOpen
  \bibfield  {author} {\bibinfo {author} {\bibfnamefont {M.~O.}\ \bibnamefont
  {Baud}}, \bibinfo {author} {\bibfnamefont {J.~K.}\ \bibnamefont {Kleen}},
  \bibinfo {author} {\bibfnamefont {E.~A.}\ \bibnamefont {Mirro}}, \bibinfo
  {author} {\bibfnamefont {J.~C.}\ \bibnamefont {Andrechak}}, \bibinfo {author}
  {\bibfnamefont {D.}~\bibnamefont {King-Stephens}}, \bibinfo {author}
  {\bibfnamefont {E.~F.}\ \bibnamefont {Chang}}, \ and\ \bibinfo {author}
  {\bibfnamefont {V.~R.}\ \bibnamefont {Rao}},\ }\bibfield  {title} {\enquote
  {\bibinfo {title} {Multi-day rhythms modulate seizure risk in epilepsy},}\
  }\href@noop {} {\bibfield  {journal} {\bibinfo  {journal} {Nat. Commun.}\
  }\textbf {\bibinfo {volume} {9}},\ \bibinfo {pages} {88} (\bibinfo {year}
  {2018})}\BibitemShut {NoStop}%
\bibitem [{\citenamefont {Karoly}\ \emph {et~al.}(2018)\citenamefont {Karoly},
  \citenamefont {Goldenholz}, \citenamefont {Freestone}, \citenamefont {Moss},
  \citenamefont {Grayden}, \citenamefont {Theodore},\ and\ \citenamefont
  {Cook}}]{karoly2018}%
  \BibitemOpen
  \bibfield  {author} {\bibinfo {author} {\bibfnamefont {P.~J.}\ \bibnamefont
  {Karoly}}, \bibinfo {author} {\bibfnamefont {D.~M.}\ \bibnamefont
  {Goldenholz}}, \bibinfo {author} {\bibfnamefont {D.~R.}\ \bibnamefont
  {Freestone}}, \bibinfo {author} {\bibfnamefont {R.~E.}\ \bibnamefont {Moss}},
  \bibinfo {author} {\bibfnamefont {D.~B.}\ \bibnamefont {Grayden}}, \bibinfo
  {author} {\bibfnamefont {W.~H.}\ \bibnamefont {Theodore}}, \ and\ \bibinfo
  {author} {\bibfnamefont {M.~J.}\ \bibnamefont {Cook}},\ }\bibfield  {title}
  {\enquote {\bibinfo {title} {Circadian and circaseptan rhythms in human
  epilepsy: a retrospective cohort study},}\ }\href@noop {} {\bibfield
  {journal} {\bibinfo  {journal} {Lancet Neurol.}\ }\textbf {\bibinfo {volume}
  {17}},\ \bibinfo {pages} {977--985} (\bibinfo {year} {2018})}\BibitemShut
  {NoStop}%
\bibitem [{\citenamefont {Takagi}(2017)}]{takagi2017sleep}%
  \BibitemOpen
  \bibfield  {author} {\bibinfo {author} {\bibfnamefont {S.}~\bibnamefont
  {Takagi}},\ }\bibfield  {title} {\enquote {\bibinfo {title} {Sleep and
  epilepsy},}\ }\href@noop {} {\bibfield  {journal} {\bibinfo  {journal} {Sleep
  Biol. Rhythms}\ }\textbf {\bibinfo {volume} {15}},\ \bibinfo {pages}
  {189--196} (\bibinfo {year} {2017})}\BibitemShut {NoStop}%
\bibitem [{\citenamefont {Khan}\ \emph {et~al.}(2018)\citenamefont {Khan},
  \citenamefont {Nobili}, \citenamefont {Khatami}, \citenamefont
  {Loddenkemper}, \citenamefont {Cajochen}, \citenamefont {Dijk},\ and\
  \citenamefont {Eriksson}}]{khan2018circadian}%
  \BibitemOpen
  \bibfield  {author} {\bibinfo {author} {\bibfnamefont {S.}~\bibnamefont
  {Khan}}, \bibinfo {author} {\bibfnamefont {L.}~\bibnamefont {Nobili}},
  \bibinfo {author} {\bibfnamefont {R.}~\bibnamefont {Khatami}}, \bibinfo
  {author} {\bibfnamefont {T.}~\bibnamefont {Loddenkemper}}, \bibinfo {author}
  {\bibfnamefont {C.}~\bibnamefont {Cajochen}}, \bibinfo {author}
  {\bibfnamefont {D.-J.}\ \bibnamefont {Dijk}}, \ and\ \bibinfo {author}
  {\bibfnamefont {S.~H.}\ \bibnamefont {Eriksson}},\ }\bibfield  {title}
  {\enquote {\bibinfo {title} {Circadian rhythm and epilepsy},}\ }\href@noop {}
  {\bibfield  {journal} {\bibinfo  {journal} {Lancet Neurol.}\ }\textbf
  {\bibinfo {volume} {17}},\ \bibinfo {pages} {1098--1108} (\bibinfo {year}
  {2018})}\BibitemShut {NoStop}%
\bibitem [{\citenamefont {Lehnertz}\ \emph {et~al.}(2014)\citenamefont
  {Lehnertz}, \citenamefont {Ansmann}, \citenamefont {Bialonski}, \citenamefont
  {Dickten}, \citenamefont {Geier},\ and\ \citenamefont {Porz}}]{lehnertz2014}%
  \BibitemOpen
  \bibfield  {author} {\bibinfo {author} {\bibfnamefont {K.}~\bibnamefont
  {Lehnertz}}, \bibinfo {author} {\bibfnamefont {G.}~\bibnamefont {Ansmann}},
  \bibinfo {author} {\bibfnamefont {S.}~\bibnamefont {Bialonski}}, \bibinfo
  {author} {\bibfnamefont {H.}~\bibnamefont {Dickten}}, \bibinfo {author}
  {\bibfnamefont {C.}~\bibnamefont {Geier}}, \ and\ \bibinfo {author}
  {\bibfnamefont {S.}~\bibnamefont {Porz}},\ }\bibfield  {title} {\enquote
  {\bibinfo {title} {Evolving networks in the human epileptic brain},}\ }\href
  {\doibase 10.1016/j.physd.2013.06.009} {\bibfield  {journal} {\bibinfo
  {journal} {Physica D}\ }\textbf {\bibinfo {volume} {267}},\ \bibinfo {pages}
  {7--15} (\bibinfo {year} {2014})}\BibitemShut {NoStop}%
\bibitem [{\citenamefont {Geier}\ \emph {et~al.}(2015)\citenamefont {Geier},
  \citenamefont {Bialonski}, \citenamefont {Elger},\ and\ \citenamefont
  {Lehnertz}}]{geier2015}%
  \BibitemOpen
  \bibfield  {author} {\bibinfo {author} {\bibfnamefont {C.}~\bibnamefont
  {Geier}}, \bibinfo {author} {\bibfnamefont {S.}~\bibnamefont {Bialonski}},
  \bibinfo {author} {\bibfnamefont {C.~E.}\ \bibnamefont {Elger}}, \ and\
  \bibinfo {author} {\bibfnamefont {K.}~\bibnamefont {Lehnertz}},\ }\bibfield
  {title} {\enquote {\bibinfo {title} {How important is the seizure onset zone
  for seizure dynamics?}}\ }\href {\doibase 10.1016/j.seizure.2014.10.013}
  {\bibfield  {journal} {\bibinfo  {journal} {Seizure}\ }\textbf {\bibinfo
  {volume} {25}},\ \bibinfo {pages} {160--166} (\bibinfo {year}
  {2015})}\BibitemShut {NoStop}%
\bibitem [{\citenamefont {Spencer}, \citenamefont {Gerrard},\ and\
  \citenamefont {Zaveri}(2018)}]{spencer2018roles}%
  \BibitemOpen
  \bibfield  {author} {\bibinfo {author} {\bibfnamefont {D.~D.}\ \bibnamefont
  {Spencer}}, \bibinfo {author} {\bibfnamefont {J.~L.}\ \bibnamefont
  {Gerrard}}, \ and\ \bibinfo {author} {\bibfnamefont {H.~P.}\ \bibnamefont
  {Zaveri}},\ }\bibfield  {title} {\enquote {\bibinfo {title} {The roles of
  surgery and technology in understanding focal epilepsy and its
  comorbidities},}\ }\href@noop {} {\bibfield  {journal} {\bibinfo  {journal}
  {Lancet Neurol.}\ }\textbf {\bibinfo {volume} {17}},\ \bibinfo {pages}
  {373--382} (\bibinfo {year} {2018})}\BibitemShut {NoStop}%
\bibitem [{\citenamefont {Ashwin}\ \emph {et~al.}(2012)\citenamefont {Ashwin},
  \citenamefont {Wieczorek}, \citenamefont {Vitolo},\ and\ \citenamefont
  {Cox}}]{ashwin2012tipping}%
  \BibitemOpen
  \bibfield  {author} {\bibinfo {author} {\bibfnamefont {P.}~\bibnamefont
  {Ashwin}}, \bibinfo {author} {\bibfnamefont {S.}~\bibnamefont {Wieczorek}},
  \bibinfo {author} {\bibfnamefont {R.}~\bibnamefont {Vitolo}}, \ and\ \bibinfo
  {author} {\bibfnamefont {P.}~\bibnamefont {Cox}},\ }\bibfield  {title}
  {\enquote {\bibinfo {title} {Tipping points in open systems: bifurcation,
  noise-induced and rate-dependent examples in the climate system},}\
  }\href@noop {} {\bibfield  {journal} {\bibinfo  {journal} {Phil. Trans. Roy.
  Soc. A: Mathematical, Physical and Engineering Sciences}\ }\textbf {\bibinfo
  {volume} {370}},\ \bibinfo {pages} {1166--1184} (\bibinfo {year}
  {2012})}\BibitemShut {NoStop}%
\bibitem [{\citenamefont {Ritchie}\ and\ \citenamefont
  {Sieber}(2016)}]{ritchie2016}%
  \BibitemOpen
  \bibfield  {author} {\bibinfo {author} {\bibfnamefont {P.}~\bibnamefont
  {Ritchie}}\ and\ \bibinfo {author} {\bibfnamefont {J.}~\bibnamefont
  {Sieber}},\ }\bibfield  {title} {\enquote {\bibinfo {title} {Early-warning
  indicators for rate-induced tipping},}\ }\href@noop {} {\bibfield  {journal}
  {\bibinfo  {journal} {Chaos}\ }\textbf {\bibinfo {volume} {26}},\ \bibinfo
  {pages} {093116} (\bibinfo {year} {2016})}\BibitemShut {NoStop}%
\bibitem [{\citenamefont {Ritchie}\ and\ \citenamefont
  {Sieber}(2017)}]{ritchie2017}%
  \BibitemOpen
  \bibfield  {author} {\bibinfo {author} {\bibfnamefont {P.}~\bibnamefont
  {Ritchie}}\ and\ \bibinfo {author} {\bibfnamefont {J.}~\bibnamefont
  {Sieber}},\ }\bibfield  {title} {\enquote {\bibinfo {title} {Probability of
  noise- and rate-induced tipping},}\ }\href@noop {} {\bibfield  {journal}
  {\bibinfo  {journal} {Phys. Rev. E}\ }\textbf {\bibinfo {volume} {95}},\
  \bibinfo {pages} {052209} (\bibinfo {year} {2017})}\BibitemShut {NoStop}%
\bibitem [{\citenamefont {McSharry}, \citenamefont {Smith},\ and\ \citenamefont
  {Tarassenko}(2003)}]{mcsharry2003prediction}%
  \BibitemOpen
  \bibfield  {author} {\bibinfo {author} {\bibfnamefont {P.~E.}\ \bibnamefont
  {McSharry}}, \bibinfo {author} {\bibfnamefont {L.~A.}\ \bibnamefont {Smith}},
  \ and\ \bibinfo {author} {\bibfnamefont {L.}~\bibnamefont {Tarassenko}},\
  }\bibfield  {title} {\enquote {\bibinfo {title} {Prediction of epileptic
  seizures: {Are} nonlinear methods relevant?}}\ }\href@noop {} {\bibfield
  {journal} {\bibinfo  {journal} {Nat. Med.}\ }\textbf {\bibinfo {volume}
  {9}},\ \bibinfo {pages} {241} (\bibinfo {year} {2003})}\BibitemShut {NoStop}%
\end{thebibliography}
\end{document}